\documentclass[3p, twocolumn,10pt,authoryear]{elsarticle}

\usepackage{amsfonts, amsmath, amsthm, amssymb}
\usepackage{ulem}
\usepackage{xcolor}
\usepackage{booktabs}
\usepackage{orcidlink}
\hypersetup{
    colorlinks=true,  
    linkcolor=blue,   
}

 \usepackage{lineno}

\newcommand{\p}{\partial}
\newcommand{\Ai}{\mathrm{Ai}}
\newcommand{\Bi}{\mathrm{Bi}}

\newcommand{\num}[1]{\times 10^{#1}}

\journal{Experimental Thermal and Fluid Science}

\begin{document}

\begin{frontmatter}

\title{Passive transverse forcing of turbulent boundary-layer flow using sinusoidal surface grooves}
\author[1]{Max W. Knoop  \corref{cor1}\orcidlink{0009-0008-8848-3006} } \ead{m.w.knoop@tudelft.nl}
\author[1]{Bas W. van Oudheusden \orcidlink{0000-0002-7255-0867} }
\author[1]{Luuk Pelkmans}
\author[1]{Ferry F.\,J. Schrijer \orcidlink{0000-0002-7532-4320}}
\affiliation[1]{organization={Faculty of Aerospace Engineering, Delft University of Technology},
addressline={Kluyverweg 1},
city={Delft},
postcode={2628 HS}, 
country={The Netherlands}}
\cortext[cor1]{Corresponding author}

\begin{abstract}
A surface geometry consisting of parallel, meandering streamwise grooves has been experimentally studied as an alternative means of passive transverse forcing of turbulent boundary-layer flow.
Particle image velocimetry (PIV) was the main diagnostic tool for characterising the properties of the induced spanwise flow.
Contrary to the original expectation, the flow does not exhibit a spanwise-uniform undulation aligned with the grooves; instead, a converging-diverging flow pattern results. 
This flow pattern can be attributed to the spanwise periodicity of the lateral pressure gradient.
The forcing effect is found to initially increase with the groove amplitude, but it saturates when the groove slope becomes too steep.
The observed induced flow, referred to as a Passive Stokes Layer (PSL), can be considered as being composed of an inertial (pressure-driven) \textit{outer solution} generated by the displacement effect of the non-smooth surface geometry, and a viscous \textit{inner solution} to accommodate the no-slip condition at the wall.
The mechanism of transverse flow generation is elucidated by an inviscid flow model that relates the forcing to the surface geometric properties, with predictions in good agreement with the experimental results.
Although a reduction in the near-wall turbulence levels over the groove surfaces is observed, no direct evidence for (mean) drag reduction is evident from the data. Instead, an estimate of the frictional drag potential is based on establishing a tentative relation to an equivalent spatial Stokes layer (SSL) induced by active wall forcing.
This theoretical comparison indicates that the induced passive forcing is sufficient to act on the (active) spanwise forcing mechanism, but produces at most a few per cent of frictional drag reduction. 
Any potential savings are likely offset by pressure drag and other losses, so that, similar to active forcing, its potential for net drag reduction in practical applications is limited.
\end{abstract}

\begin{graphicalabstract}
\includegraphics[width = \textwidth]{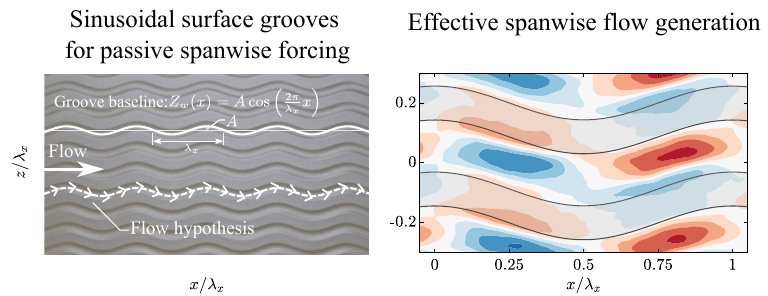}
\end{graphicalabstract}

\begin{highlights}
\item Passive grooves effectively generate a Stokes-like spanwise flow in turbulence.
\item Surface shape produces pressure gradients that cause an inertial forcing mechanism.
\item Experiments and theory show a passive Stokes layer with viscous and inertial regions.
\item Viscous-layer strain is sufficient to act on the active spanwise-forcing mechanism. 
\item Practical applicability may however be limited by losses such as pressure drag.
\end{highlights}


\end{frontmatter}


\section{Introduction}\label{sec:intro}

Flow control aimed at reducing turbulent skin-friction drag can lead to substantial energy savings, which may, for example, yield reduced carbon dioxide emissions in practical applications.
A particularly effective strategy to achieve this type of drag reduction (DR) is the active and pre-determined imposition of a spatio-temporal wave of spanwise wall motion, for which high DR values of over 45\% have been reported \citep[for a review see][]{ricco_review_2021}.
This spanwise forcing induces a thin transverse-velocity shear layer -- referred to as the Stokes layer \citep{quadrio_laminar_2011} -- that periodically interacts with the near-wall turbulence, driving the underlying DR mechanism \citep{choi1998turbulent, choi2001mechanism, ricco_effects_2004, Touber_near-wall_2012, Agostini_spanwise_2014, agostini_turbulence_2015, knoop2025response}. 
While theoretical net-power savings of over 25\% have been reported \citep{quadrio_streamwise-travelling_2009}, in practice the overall power expenditure required to drive the spanwise flow in combination with losses in the entire actuation chain vastly outweighs these savings, by several orders of magnitude \citep[see, e.g.,][]{auteri2010experimental,gatti2015experimental,bird2018experimental}. In addition, these active techniques are not easily adaptable to practical implementation in view of the system complexity associated with active forcing, which limits their adoption despite their promise of high DR and energy savings.\\

These considerations have motivated the study of passive techniques that aim to induce a spanwise flow actuation similar to that induced by active techniques.
More specifically, the active analogue that such passive forcing methods aim to recreate is the steady spatially periodic forcing 
that imposes a spanwise wall velocity $w_w(x)$, according to
\begin{equation}
    \label{eq:w_wSSL}
    w_w(x) = W_\text{SSL}\sin(k_x x) , 
\end{equation}
where $W_\text{SSL}$ is the spanwise velocity amplitude, and $k_x = {2\pi}/{\lambda_x}$ indicates the streamwise wavenumber/wavelength. Throughout this paper, capital $W$ is used to indicate the transverse velocity amplitude.
This active forcing induces a so-called spatial Stokes layer (SSL), for which an analytical solution was formulated in \citet{viotti_streamwise_2009}.
Instead of wall-based forcing, these passive techniques produce a forcing of the bulk flow overlying the static wall.
The efficacy of bulk-flow forcing is apparent from active strategies, based on, e.g., plasma actuators \citep{wilkinson2003investigation,corke2018active,thomas2019turbulent}, acoustic excitation \citep{Schaafsma2025transverse}, and Lorentz force \citep{berger2000turbulent,lee2002control,breuer2004actuation}, that report similar DR to that of wall-based forcing.

In this study, we focus on passive strategies based on large-scale geometrical modifications of the surface, which include existing techniques such as oblique wavy walls \citep{chernyshenko2013drag, ghebali_large-scale_2017} and shallow spherical dimples \citep{lienhart2008drag, tay2015mechanics, Nesselrooij_drag_2016, vanCampenhout2023experimental}.
Both techniques effectively generate a cross-flow component by introducing a three-dimensional pressure gradient.

\begin{figure}
    \centering
    \includegraphics[width = \columnwidth]{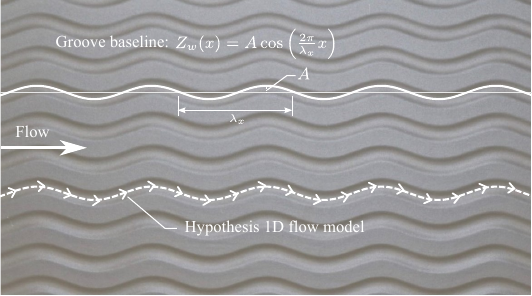}
    \caption{Photograph of a SU test surface. Overlayed are the groove baseline definition and the hypothesised working principle (refer to \S\ref{sec:intro-model-1D}).}
    \label{fig:flowHypothesis}
\end{figure}

The present investigation addresses an alternative surface geometry described in a patent by \citet{vanNesselrooij2020body}, and which is referred to as sinusoidal undulations (SU).
A top-view of an SU test surface is shown in figure~\ref{fig:flowHypothesis}.
The test-surface geometry constitutes shallow surface grooves that meander along the streamwise direction, with the proposed working principle that this will likewise induce an oscillatory spanwise component in the near-wall flow, in direct analogy to the active spatial (standing-wave) actuation referred to above. 
While SU may appear to resemble sinusoidal riblets, investigated for the same purpose of frictional DR \citep{peet2008turbulent,sasamori2014experimental,cafiero2022drag,cafiero2024manipulation}, a key difference is their large-scale and shallow surface geometry versus the microscopic scale of riblets that make them sensitive to erosion and fouling in practice.

\subsection{One-dimensional model for the transverse flow}
\label{sec:intro-model-1D}

A simplified prediction of the induced spanwise velocity is made by assuming that the flow is forced to follow the SU geometry perfectly. 
This hypothesis is visualised in figure~\ref{fig:flowHypothesis}.
The SUs are described by a sinusoidal baseline
\begin{equation}
    \label{eq:grooveBaseline}
    Z_w(x) = A \cos (k_x x) = A \cos\left(\frac{2\pi}{\lambda_x}x\right), 
\end{equation}
where $x$ denotes the streamwise coordinate, $A$ is the groove spanwise displacement amplitude, and $k_x$ and $\lambda_x$ are the streamwise wavenumber and wavelength. Assuming a constant convective velocity $U_c$ provides a one-dimensional (1D) prediction of the spanwise velocity, as

\begin{equation}
    \label{eq:1Dmodel}
    w(x)= \frac{d Z_w}{d t}=\frac{d x}{d t} \frac{d Z_w}{d x} = 
    -\underbrace{A k_x U_c }_{W_\text{1D}} \sin (k_x x),
\end{equation}
where $W_\text{1D}$ is the predicted spanwise velocity amplitude.
For turbulent boundary layers, the near-wall convection velocity is $U_c \approx 10 U_\tau$ \citep{kim1993propagation}, where $U_\tau$ is the skin-friction velocity. Based on this, the 1D model predicts $W_\text{1D}/U_\tau \sim 3$ to 8 for the present SU geometries (refer to table~\ref{tab:modelGeometry}), which falls in the range of typical actively imposed velocity amplitudes of $2$ to 12 $U_\tau$. 

This conceptual model is evidently a significant oversimplification of the real flow behaviour. 
Firstly, given the discrete groove arrangement, the flow is likely to exhibit spanwise variation rather than a spanwise uniform deflection pattern. 
Moreover, the model does not take into account that the efficiency of the spanwise forcing is likely dependent on the groove cross-sectional geometry, if only for the reason that a zero groove depth will not result in any forcing. 
Hence, the forcing efficiency is expected to increase with groove depth. \\

Later in this paper, we will consider a more realistic flow model that relates the lateral forcing mechanism to the groove geometry, providing a quantitative illustration and prediction of the flow forcing performance, and serving as a basis for comparison with experimental observations.

\subsection{Research objectives}\label{sec:intro-contributions}
The research question this study aims to address is: \textit{Can sinusoidal undulations induce effective passive forcing with potential for turbulent drag reduction?}

Five SU test geometries were investigated in a turbulent boundary layer (TBL) flow to establish the effect of varying groove amplitude $A$ and cross-sectional geometry (details in \S\ref{sec:method-testsurfaces}). 
To characterise the flow over the test surfaces, particle image velocimetry measurements (details in \S\ref{sec:method-PIV}) were conducted in the streamwise-wall-normal plane and in a near-wall wall-parallel plane.
Results show that a passive type Stokes layer -- referred to as the \textit{passive Stokes layer} (PSL) -- is induced by the test surfaces; its overall flow organisation for the mean flow and turbulence is established in \S\ref{sec:flow-organisation}.
We elucidate the mechanism of spanwise flow generation in \S\ref{sec:mechanism}, supported by a potential-flow model that is detailed in \S\ref{app:potential}.
The spanwise forcing characteristics of the PSL are quantified in \S\ref{sec:forcing-characteristics} and compared with a predictive scaling derived from the SU geometric parameters using the potential-flow model. 
To draw an analogy with active spanwise wall-forcing, a three-dimensional (3D) viscous model of the PSL, derived in \ref{app:PSL}, is used in \S\ref{sec:DR-discussion} to compare with the active spatial Stokes layer (SSL) and to evaluate the performance; discussion on the DR potential is provided accordingly.

\section{Methodology}\label{sec:method}

\subsection{Facility and test conditions}\label{sec:method-facility}
\begin{figure*}
    \centering
    \includegraphics[width = \textwidth]{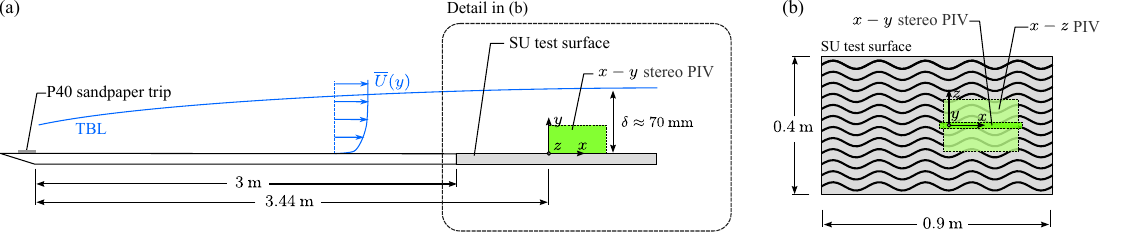}
    \caption{(a) Schematic of wind-tunnel configuration and (b) detail of the particle image velocmetry (PIV) experiments over the SU test-surface.}
    \label{fig:expSetup}
\end{figure*}

Experiments have been conducted in an open-return wind tunnel with a cross-sectional area of $0.4\times 0.4$\:m$^2$ and a typical free-stream turbulence intensity of 0.7\%. Figure~\ref{fig:expSetup}(a) provides a schematic of the experimental setup for which the wind-tunnel configuration matches the experiments in \citet{carrasco2024experimental}. 
A boundary layer was generated on a 30-mm thick flat plate with an elliptic leading edge and tripped to turbulence by a strip of 24-grit sand-grain roughness.
A canonical flat-plate TBL subsequently developed over a streamwise extent of 3\:m, after which the test surfaces under investigation were mounted flush with the wall of the incoming TBL.

Viscous scaling, indicated by the superscript `$+$', is adopted using the kinematic viscosity $\nu$ and the skin-friction velocity $U_\tau$ over a flat-plate reference surface. The characteristic viscous length-scale is defined by $\delta_v = \nu/U_\tau$. 
Table~\ref{tab:conditions} shows the experimental conditions and viscous scaling parameters.
Experiments were carried out for free-stream velocities of $U_\infty = 5$ and 7.5\:m/s, with corresponding $Re_{\tau} = \delta U_\tau/\nu$ of 1020 and 1410, where $\delta$ is the boundary layer thickness. 

The streamwise, wall-normal, and spanwise coordinates are denoted by $(x,y,z)$, and the corresponding instantaneous velocity components are $(u,v,w)$; $p$ denotes pressure. Instantaneous quantities are decomposed into a mean and fluctuating part as denoted by an overbar and prime superscript, e.g., $u = \overline{u} + u'$. The $y$ coordinate starts at the top face of the SU geometry (the flat wall regions in between the grooves).
As indicated in figure~\ref{fig:expSetup}, the origin of the coordinate system is located at the upstream edge of the PIV field-of-view, which is $2.44$\:m downstream of the trip.

\begin{table}[]
\centering
\begin{tabular}{lllll}
$U_\infty$ (m/s) & $\delta$ (mm) & $U_\tau$ (m/s) & $\delta_v$ (\textmu m) & $Re_\tau$ \\ \midrule
5 & 70 & 0.217 & 68.9 & 1020 \\
7.5 & 69 & 0.308 & 48.6 & 1410
\end{tabular}
\caption{Overview of the free-stream conditions, boundary layer characteristics, and viscous scaling parameters}
\label{tab:conditions}
\end{table}

\begin{figure*}
    \centering
    \includegraphics[width = \textwidth]{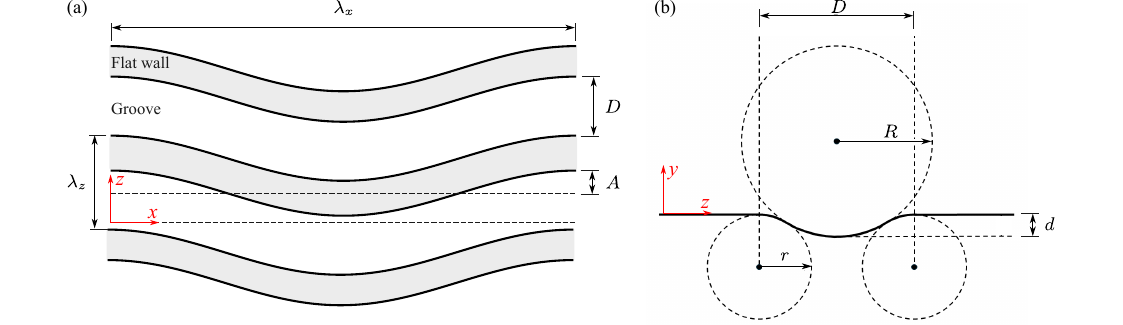}
    \caption{Schematic of the SU geometry, (a) top-view, (b) groove cross-section.}
    \label{fig:geometry}
\end{figure*}

\subsection{Test surfaces}\label{sec:method-testsurfaces}
A flat-plate reference surface and five SU models are considered. 
Each test surface was CNC-machined and measures $881 \times 366$\:mm$^2$ in the streamwise and spanwise direction.
The surfaces were wrapped with a vinyl sticker to smooth machining marks (typically less than 0.01 mm) and provide a consistent surface finish across all test surfaces.

The SU geometry, schematically shown in top-view in figure~\ref{fig:geometry}(a), consists of grooves characterised by the sinusoidal baseline defined in equation \eqref{eq:grooveBaseline} with streamwise wavelength $\lambda_x$ and spanwise displacement amplitude $A$, while the spacing of adjacent grooves is given by its spanwise wavelength $\lambda_z$.
The cross-sectional groove profile shown in figure~\ref {fig:geometry}(b) is derived from the dimple geometry in \citet{Nesselrooij_drag_2016} and is defined by its width $D$ and depth $d$, with an edge rounding radius $r = D$.

\begin{table*}[h]
\centering
\begin{tabular}{llllllllll}
 & \multicolumn{5}{c}{Dimensional (mm)} & \multicolumn{4}{c}{Scaled} \\
 \cmidrule(lr){2-6} \cmidrule(lr){7-10} \\ 
Model ID & $A$  & $D$  & $d$  & $\lambda_x$  & $\lambda_z$ & $\alpha_{\max}$ ($^\circ$) & $A/\lambda_x$ & $d/D$ & $\lambda_z/D$ \\
\midrule
A3-D10-d050 & 3.24 & 10 & 0.5 & 57.18 & 16.5 & 19.6 & 0.057 & 0.050 & 1.65 \\
A5-D10-d050  & 5.25 & 10 & 0.5 & 57.18 & 16.5 & 30.0 & 0.092 & 0.050 & 1.65 \\
A7-D10-d050 & 7.64 & 10 & 0.5 & 57.18 & 16.5 & 40.0 & 0.134 & 0.050 & 1.65 \\
A3-D10-d025 & 3.24 & 10 & 0.25 & 57.18 & 16.5 & 19.6 & 0.057 & 0.025 & 1.65 \\
A3-D20-d100 & 3.24 & 20 & 1 & 57.18 & 33 & 19.6 & 0.057 & 0.050 & 1.65
\end{tabular}
\caption{Geometrical parameters of the SU test surfaces, the left side reports the dimensional values in mm, the right side reports the appropriate scaling of the parameters.}
\label{tab:modelGeometry}
\end{table*}

\begin{table*}[h]
\centering
\begin{tabular}{lllllllllll}
 & \multicolumn{5}{c}{$U_\infty = 5$ m/s} & \multicolumn{5}{c}{$U_\infty = 7.5$ m/s} \\
\cmidrule(lr){2-6} \cmidrule(lr){7-11} \\ 
Model ID & $A^+$ & $D^+$ & $d^+$ & $\lambda_x^+$ & $\lambda_z^+$ & $A^+$ & $D^+$ & $d^+$ & $\lambda_x^+$ & $\lambda_z^+$ \\ \midrule
A3-D10-d050 & 47.9 & 145 & 7.3 & 830 & 240 & 66.7 & 206 & 10.3 & 1180 & 340 \\
A5-D10-d050 & 76.2 & 145 & 7.3 & 830 & 240 & 108.1 & 206 & 10.3 & 1180 & 340 \\
A7-D10-d050& 110.9 & 145 & 7.3 & 830 & 240 & 157.3 & 206 & 10.3 & 1180 & 340 \\
A3-D10-d025 & 47.9 & 145 & 3.6 & 830 & 240 & 66.7 & 206 & 5.1 & 1180 & 340 \\
A3-D20-d100 & 47.9 & 290 & 14.5 & 830 & 479 & 66.7 & 412 & 20.6 & 1180 & 680
\end{tabular}
\caption{Viscous scaled geometric parameters of the SU test surfaces.}
\label{tab:modelGeometryViscous}
\end{table*}

Table~\ref{tab:modelGeometry} presents an overview of the five test surfaces considered to assess the effect of groove amplitude and cross-sectional shape. 
Model A3-D10-d05 serves as a baseline geometry, designed in analogy to the staggered dimple configuration `plate A' in \citet{Nesselrooij_drag_2016} by setting $\lambda_x$ and $\lambda_z$ to the streamwise and spanwise dimple spacing and matching $d/D$.
Our results show that the groove baseline is appropriately scaled by $A/\lambda_x$ while the spanwise cross-section is appropriately scaled using $D$, i.e., $d/D$ and $\lambda_z/D$.
The maximum slope of the grooves, which occurs at $x/\lambda_x = 0.25,0.75$, provides an indicator of the spanwise velocity amplitude, based on the 1D model (refer to \S\ref{sec:intro-model-1D}), and is given by
\begin{equation}
    \label{eq:alphaMax}
    \alpha_{\max} = \tan^{-1}(Ak_x).
\end{equation}

The viscous scaled geometric parameters are presented in table~\ref{tab:modelGeometryViscous}, which shows that the streamwise wavelength is close to $\lambda_x = 1000$ where optimum DR for active forcing occurs \citep{viotti_streamwise_2009, knoop2025response}.

\subsection{Particle image velocimetry}\label{sec:method-PIV}
Particle image velocimetry (PIV) experiments were conducted in two different plane orientations, as shown in figure~\ref{fig:expSetup}(b).
In both experiments, one or two digital LaVision sCMOS cameras ($2160 \times 2560$ pixels, 16-bit, 6.5 \textmu m pixel size) were used for image acquisition, an Evergreen Nd:YAG (532 nm) 200 mJ/pulse laser was used for illumination, and 1 \textmu m water-glycol droplets were used for seeding. 

Planar two-velocity component (2D-2C) measurements were conducted in a wall-parallel $(x,z)$ plane with a field-of-view of  $92\:\times\:78$\:mm$^2$. One sCMOS camera was used with a 105-mm focal-length lens at aperture $f/5.6$.  
The laser sheet was centred at $y \approx 1$\:mm, with a $\Delta y_s \approx 1.4$\:mm thickness, i.e., the illumination covered a region of $y_s\approx 1\pm 0.7$\:mm.
In viscous units, this corresponds to a laser sheet located at $y^+_s\approx 15\pm 10$ and $21\pm 14$ for $U_\infty$ = 5.0 and 7.5\:m/s, respectively.
For image acquisition, 2000 two-pulse snapshots were acquired at 15\:Hz, with a time separation of $\Delta t = 150$ and 100\:\textmu s between image pairs for the two respective free-stream velocities.
To remove background laser-sheet reflections in the raw images, each image was divided by the time-average over all images. 
Velocity vectors were computed using a multi-pass (3 passes) cross-correlation algorithm with an iterative window deformation method \citep{scarano2000advances,scarano2002iterative}. The final pass used circular interrogation windows (IWs) of 48\:$\times$\:48 pixels, at an overlap factor of 75\%; this choice resulted in a viscous-scaled spatial resolution (i.e., IW size) of approximately $25\:\times\:25\ \delta_v^2$ and $35\:\times\:35\ \delta_v^2$, respectively.

In addition, stereoscopic-PIV (2D-3C) measurements were conducted in the streamwise-wall-normal $(x,y)$ plane aligned along the undulation centerline.
The field-of-view spanned $67\:\times\:52$\:mm$^2$, with a laser-sheet thickness of $\Delta z_s \approx 1.3$\:mm.
For these measurements, two cameras were used with the same 105\:mm focal length and an aperture of $f/16$. The cameras were calibrated using a pinhole model with a LaVision type-7 calibration plate (5-mm marker spacing). 
Ensembles of 1000 two-pulse snapshots were acquired at 6\:Hz and with $\Delta t = 105$ and 75 \textmu s. 
The same image pre-processing procedure as for the planar measurements was adopted. To enhance the wall-normal resolution, elliptical (4:1) IWs were used for the vector calculations. The final IW size was 48\:$\times$\:12 pixels at 75\% overlap, resulting in a viscous spatial resolution of approximately $25\:\times\:6\ \delta_v^2$ and $35\:\times\:9\ \delta_v^2$.

\begin{figure*}[t]
    \centering
    \includegraphics[width = \textwidth]{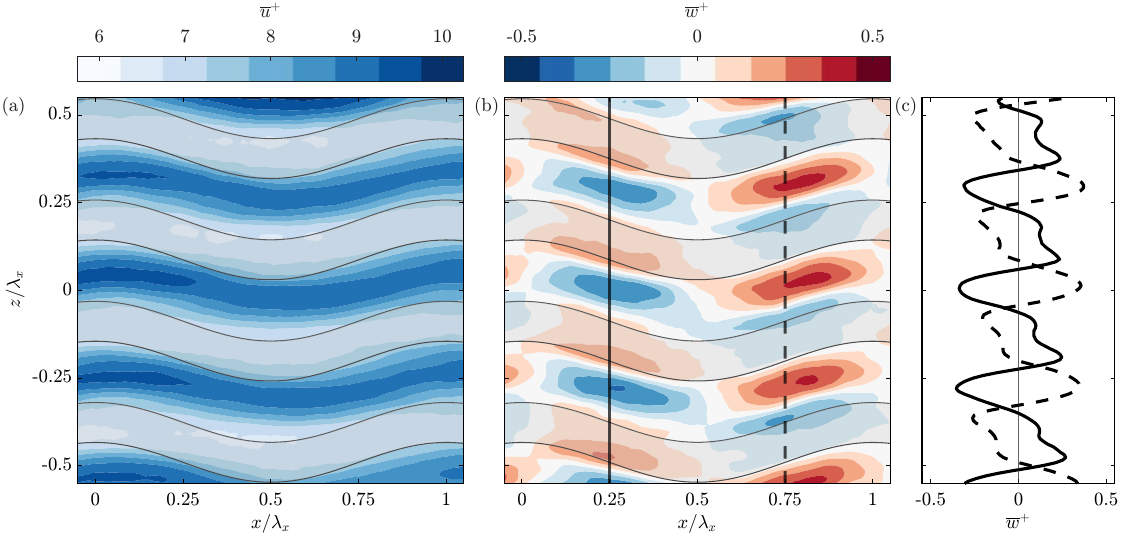}
    \caption{Mean flow in the wall-parallel plane at $y_s^+ \approx 15 \pm 10$ for SU model A3-D10-d050 at $Re_\tau = 1020$. (a) Streamwise velocity $\overline{u}^+$, (b) spanwise velocity $\overline{w}^+$, and (c) spanwise $\overline{w}^+$ profiles located at $x/\lambda_x = [0.25, 0.75]$ as indicated by the vertical lines in (b) with the same linestyle. Grey-shaded regions represent the flat wall between grooves; grooves are unshaded.}
    \label{fig:xzMeanFlow}
\end{figure*}

\section{Overall flow organisation}\label{sec:flow-organisation}
Model A3-D10-d050 is considered as the baseline geometry in the current results discussion, for which the mean flow is characterised in the near-wall wall-parallel plane in \S\ref{sec:flow-organisation-xz} and in the streamwise-wall-normal plane in \S\ref{sec:flow-organisation-xy}. The effect of the SU test surfaces on the near-wall turbulence is described in \S\ref{sec:flow-organisation-turbulence}. All presented results correspond to the measurements at $U_\infty = 5$\:m/s and $Re_\tau = 1020$ (details in table~\ref{tab:conditions}).

\subsection{Near-surface wall-parallel mean flow}\label{sec:flow-organisation-xz}

\begin{figure*}[t]
    \centering
    \includegraphics[width = \textwidth]{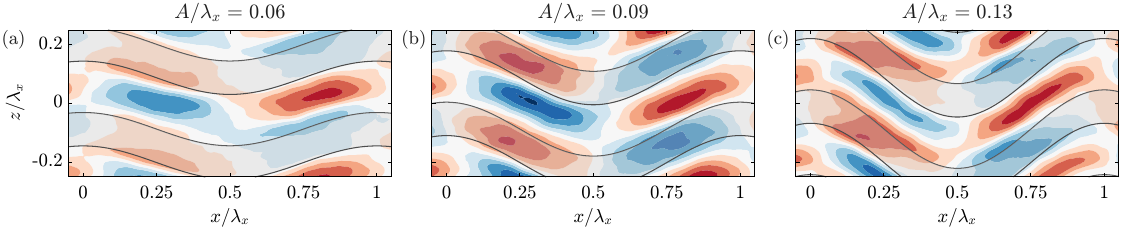}
    \caption{Effect of increasing groove amplitude $A$ on $\overline{w}^+$ in the wall-parallel plane for SU models (a) A3-D10-d050, (b) A5-D10-d050, and (c) A7-D10-d050. Colour scales are the same as in figure~\ref{fig:xzMeanFlow}.}
    \label{fig:zAmplitudeTopology}
\end{figure*}

The wall-parallel flow organisation, corresponding to the region $y^+ \approx 15 \pm 10$, is characterised in figure~\ref{fig:xzMeanFlow} in terms of the streamwise (figure~\ref {fig:xzMeanFlow}a) and spanwise (figure~\ref {fig:xzMeanFlow}b) velocity component fields.
Spanwise $\overline{w}$ profiles at $x/\lambda_x = 0.25$ and 0.75, which correspond to the locations where the groove slope is steepest (i.e., $\alpha_{\max}$ in \eqref{eq:alphaMax}), are shown in figure~\ref {fig:xzMeanFlow}(c). 
In this top-view of the flow, the transparent grey shading indicates the flat wall regions between the grooves, whereas the grooves are unshaded.

The streamwise flow is modulated by the surface, with regions of high and low streamwise velocity that coincide, respectively, with the grooves ($\overline{u}^+ \approx 9$) and the flat wall between them ($\overline{u}^+ \approx 6.5$).
Since the grooves lie below the top face of the test surface, measurements over the grooves are located farther from the surface within the overlying boundary layer, and consequently, the velocity is higher there. 
These results appear to confirm that the near-wall boundary layer follows the surface geometry and remains predominantly attached.

The spanwise velocity in figure~\ref {fig:xzMeanFlow}(b) follows the grooves and qualitatively exhibits the hypothesised behaviour based on the 1D model in \S\ref{sec:intro-model-1D}.
Namely, the strongest transverse flow is induced where the groove slope is steep and thus scales in proportion to and matches the sign of $\mathrm{d}W/\mathrm{d}x$ in equation \eqref{eq:1Dmodel}.
The primary flow regions that follow the groove are alternated in the spanwise direction by patterns with opposing values in the regions between the grooves (transparent grey shading).
This feature was not predicted by the 1D model and indicates a secondary flow directed across the groove edges rather than along its axis.
The primary and secondary flows are driven by a spanwise-periodic pressure gradient induced by the groove geometry; its mechanism is further elucidated in \S\ref{sec:mechanism}.

Inside the grooves, the primary flow creates the strongest spanwise flow of $\overline{w}^+ \approx \pm 0.4$, which occurs right downstream of the maximum groove slope at $x/\lambda_x = 0.3$ and 0.8.
Based on the profiles in figure~\ref {fig:xzMeanFlow}(c), the secondary flow has a lower magnitude with double peaks.
These peaks align with the groove edges, and the largest of the two secondary peaks occurs where the directions of the primary and secondary flows diverge.

Figure~\ref{fig:zAmplitudeTopology} shows the effect of increasing the groove amplitude $A$ on the spanwise flow, where $A/\lambda_x = 0.06$ corresponds to the results discussed above.
As the amplitude increases from $A/\lambda_x = 0.06$ to 0.09 in figures~\ref{fig:zAmplitudeTopology}(a) and (b), the $\overline{w}$ magnitude increases as may be expected based on a proportional scaling according to the 1D model (refer to equation \eqref{eq:1Dmodel}).
This effect, however, is observed to saturate for the largest amplitude in figure~\ref{fig:zAmplitudeTopology}(c) and does not mark a further increase in spanwise velocity. 
For $A/\lambda_x = 0.13$ it appears that the groove meandering is too strong to effectively contain the near-wall flow, which instead spills over the sides of the grooves.

\subsection{Streamwise-wall-normal mean flow}\label{sec:flow-organisation-xy}
\begin{figure*}
    \centering
    \includegraphics[width = \textwidth]{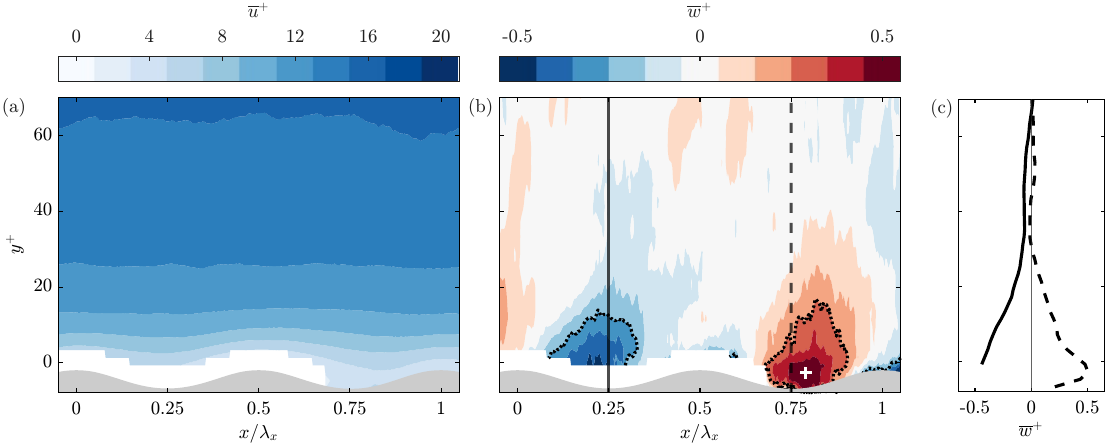}
    \caption{Mean flow in the streamwise-wall-normal plane along the centerline of a groove for SU model A3-D10-d050 at $Re_\tau = 1020$. (a) Streamwise velocity $\overline{u}^+$, (b) spanwise velocity $\overline{w}^+$, and (c) wall-normal $\overline{w}^+$ profiles located at $x/\lambda_x = [0.25, 0.75]$ as indicated by the vertical lines in (b) with the same linestyle. The grey region represents the SU surface.}
    \label{fig:xyMeanFlow}
\end{figure*}

Figure~\ref{fig:xyMeanFlow} shows the mean flow in a streamwise-wall-normal plane along the centerline of the groove, i.e., at $z = 0$. The wall is indicated by the grey patch. Note that for $x/\lambda_x<0.6$, near-wall data below $\hat y^+ \approx 5$ were removed due to limited optical access to and laser-sheet reflections on the inside of the groove. 
To account for the height variation of the test surfaces, a local wall-normal distance $\widehat y$ is introduced that is measured from the actual surface.
Figure~\ref{fig:xyMeanFlow}(a) shows that the $\overline{u}$ isolines deflect towards the surface for $y^+<20$. 
No reverse streamwise flow occurs, confirming that the flow remains attached to these shallow surface grooves (note that the data aspect ratio in the figure is approximately 10:1 in $x$:$y$).   
For the spanwise flow in figure~\ref{fig:xyMeanFlow}(b), a streamwise-periodic alternating flow appears that shows resemblance to the characteristics of the active spatial Stokes layer (SSL) in \citet{viotti_streamwise_2009}.

There are however key differences between the SSL and the present spanwise flow. 
In the active SSL situation, the spanwise velocity amplitude $W_\text{SSL}$ is imposed at the wall, whereas here the spanwise flow is constrained to $\overline{w} = 0$ at the surface in view of the no-slip condition. 
Consequently, the spanwise flow is bulk-driven by the 3D pressure gradient induced by the surface, such that $W_{\max}$ occurs away from the wall. 
The profiles in figure~\ref{fig:xyMeanFlow}(c) show that a type of \textit{two-region shear layer} forms with an inflexion around the location of $W_{\max}$, similar to that of a wall jet \citep{launder1983turbulent}. 
A thin, viscous inner layer forms, which is most alike the SSL, while the outer layer exhibits a more gradual wall-normal decay, which is characterised by an outer penetration depth $\Delta$. We refer to this shear layer as the \textit{passive Stokes layer} (PSL).

We find that for the considered case, $W_{\max}^+ \approx 0.65$ is located at $\hat y^+ \approx 6$ away from the surface, indicated by the white-plus marker in figure~\ref{fig:xyMeanFlow}(b). 
This height is similar to the inner-layer penetration depth for the SSL at equivalent $\lambda_x$.
To find the penetration depth of the outer layer, we take the contour where $|\overline{w}|$ decays to $W_{\max}e^{-1}$, shown by black-dotted isoline in figure~\ref{fig:xyMeanFlow}(b), and select the highest $y$-location to be $\Delta$, defined as
\begin{equation}
    \Delta := \max \left\{\, y \;\middle|\; \overline{w}(x,y)=W_{\max}e^{-1} \right\}.
\end{equation}
In figure~\ref{fig:xyMeanFlow}(b), $\Delta^+ \approx 20$, which is indeed much thicker than the viscous inner layer.
This deeper outer layer is caused by the wall-normal decay of the near-wall pressure gradients induced by the surface, and is of inertial rather than viscous nature.

\subsection{Effects on turbulence}\label{sec:flow-organisation-turbulence}

\begin{figure}[t]
    \centering
    \includegraphics[width = \columnwidth]{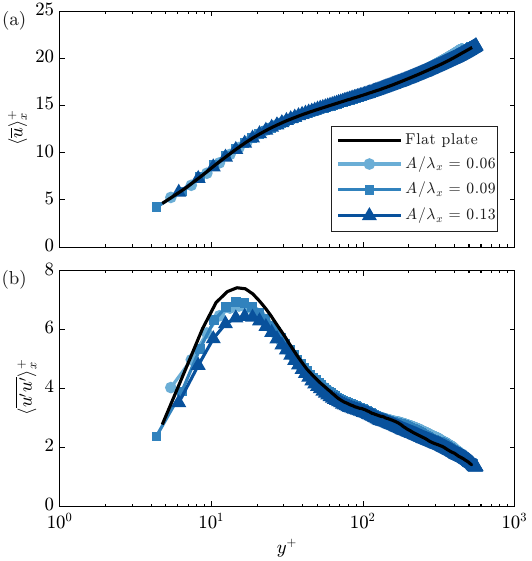}
    \caption{Wall-normal turbulence statistics profiles for increasing groove amplitude at $Re_\tau = 1020$. (a) Streamwise velocity $\langle \overline{u}\rangle^+_x$ and (b) streamwise normal stress $\langle \overline{u'u'}\rangle^+_x$.}
    \label{fig:statProfiles}
\end{figure}

For the three groove amplitudes depicted in figure~\ref{fig:zAmplitudeTopology}, wall-normal profiles of the mean streamwise velocity $\langle{\overline{u}}\rangle _x$, and the streamwise normal Reynolds stress $\langle{\overline{u'u'}}\rangle _x$ are shown in figure~\ref{fig:statProfiles}. 
The $\langle ...\rangle_x$ operator denotes streamwise averaging of the statistics across $0\leq x\leq \lambda_x$.
Measurements over a flat-plate reference surface are also included, and scaling is based on $U_\tau$ of the flat-plate TBL to appreciate the absolute changes in the statistics.

The streamwise normal stress in figure~\ref{fig:statProfiles}(b) is representative of the near-wall high- and low-speed velocity streaks that are central to the self-sustaining nature of turbulence \citep{kline1967structure, jimenez1999autonomous}.
Compared with the flat-plate reference (in black), the SU test surfaces display a sizable reduction in the inner peak at $y^+ = 15$. 
For the two lower amplitudes $A/\lambda_x = 0.06$ and 0.09, the reduction is similar, even though the latter produces a higher $\overline{w}$ magnitude in the wall-parallel plane (refer to figure~\ref{fig:zAmplitudeTopology}). 
The strongest effect is produced by the large-amplitude surface $A/\lambda_x = 0.13$ that shows a 13\% reduction of inner-peak relative to the flat-plate reference.   
While these results appear to suggest that the near-wall streaks are suppressed to a certain extent, the mean streamwise velocity figure~\ref{fig:statProfiles}(a) does not exhibit discernible differences from the flat-plate TBL or signs of a drag-reduced flow, such as the near-wall ($y^+<20$) momentum deficit that occurs for DR by active spanwise forcing.  

\begin{figure}[tbp]
    \centering
    \includegraphics[width = \columnwidth]{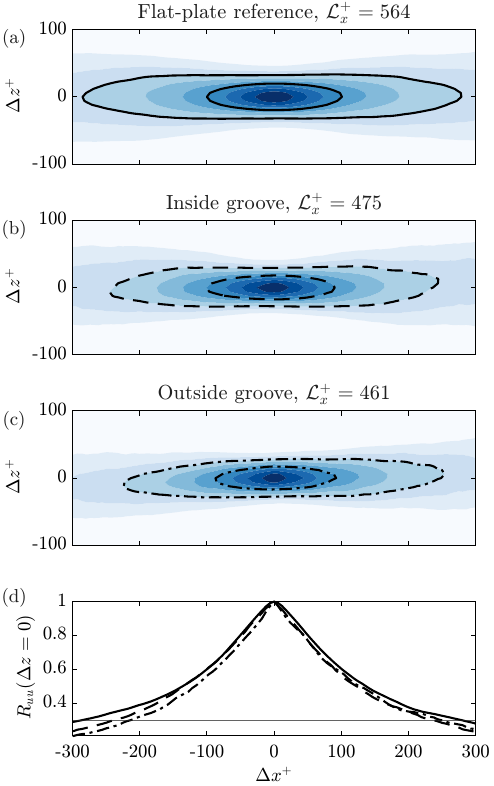}
    \caption{Two-point correlation of the streamwise velocity $R_{uu}$ for (a) flat-plate reference, and for SU model A3-D10-d050 located (b) inside the groove and (c) outside the groove. Contour levels range [0:0.1:1] and black lines denote reference levels $R_{uu} = [0.3, 0.6]$. 
    (d) Streamwise profiles at $z_0$; linestyles correspond to the reference contour lines in (a-c). 
    $R_{uu}$ was computed at $x_0/\lambda_x = 0.25$, and ensemble averaged for (b) $z_0/\lambda_z = [-1, 0, 1, 2]$ and (c) $z_0/\lambda_z = [-1.5, -0.5, 0.5, 1.5]$.}
    \label{fig:Ruu}
\end{figure}

No significant spatial variation of $\overline{u'u'}$ occurred in the wall-parallel plane (not shown), likely owing to the near-wall structures being comparable in size (approximately $1000\times 100\ \delta_v^2$ in $x$ and $z$) to the wavelengths of the test surfaces, i.e., $\lambda_x^+ = 830$ by $\lambda_z^+ = 240$.
Instead, the streamwise-spanwise coherence of the near-wall structures is investigated using the two-point correlation coefficient, defined by
\begin{equation}
    \label{eq:Ruu}
    R_{uu} = \frac{\overline{ u'(x_0,z_0) u'(x_0 + \Delta x,z_0 + \Delta z) }}{\overline{u'u'}(x_0,z_0)}.
\end{equation}
Figure~\ref{fig:Ruu} shows $R_{uu}$ for a flat-plate reference and the $A/\lambda_x = 0.06$ baseline SU test surface discussed in sections \S\ref{sec:flow-organisation-xz} and \S\ref{sec:flow-organisation-xy}. 
$R_{uu}$ is computed along $x_0/\lambda_x = 0.25$, where the groove slope is steepest, and for two spanwise locations: \textit{inside the groove} on its central axis and \textit{outside the groove} centred on the flat-wall in between the grooves.
A streamwise coherence length-scale $\mathcal{L}_x$ is defined by the width of the reference contour $R_{uu} = 0.3$ in figures~\ref{fig:Ruu}(a-c). 
The reference level is also indicated by the horizontal line in figure~\ref{fig:Ruu}(d) of the streamwise $R_{uu}(z_0)$ profiles along the spanwise centerline.

Both inside and outside the groove, in figures~\ref{fig:Ruu}(b) and (c), the streamwise coherence is shortened relative to the flat-plate reference, figure~\ref{fig:Ruu}(a).
This shortening is most notable for reference contour level $R_{uu}=0.3$, while $R_{uu}=0.6$ remains similar, as also confirmed in figure~\ref{fig:Ruu}(d).
The structure seems to be most affected by the transverse flow outside the grooves where the secondary flow occurs, for which $\mathcal{L}_x$ is reduced by 18\%; the spanwise coherence is also notably narrower here. 
At this location, $R_{uu}$ shows a slight inclination that may be caused by titling by the mean flow, i.e., the upward incline matches the sign of $\overline{w}>0$. 

\section{Mechanisms of transverse flow generation}\label{sec:mechanism}
\begin{figure*}
    \centering
    \includegraphics[width = \textwidth]{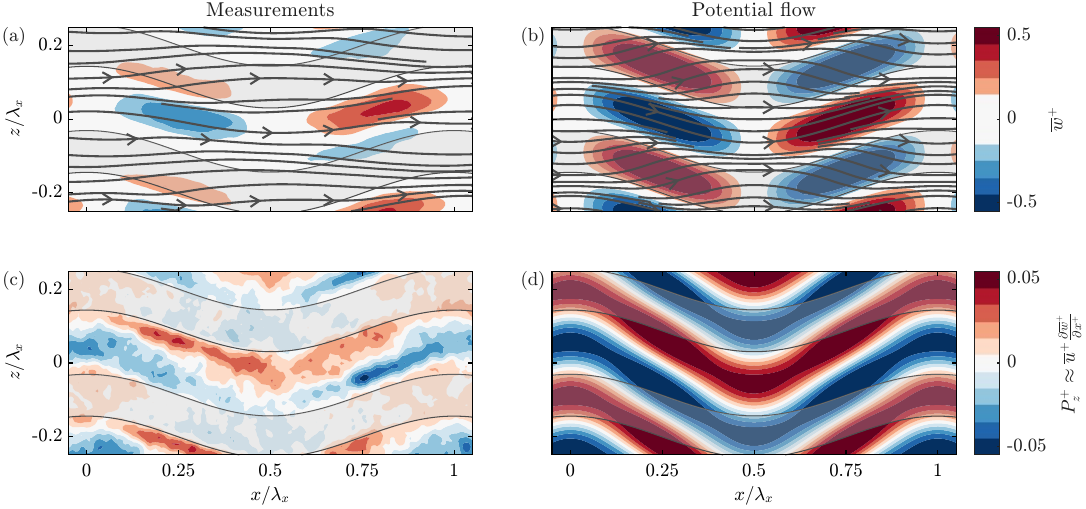}
    \caption{Comparison of wall-parallel flow for (a,c) measurements over A3-D10-d050 and (b,d) potential flow. (a,b) Mean spanwise velocity $\overline{w}^+$ and streamlines for which spanwise velocity is $5\overline{w}$ to support interpretability. (c,d) Streamwise advection term $\overline{u}^+ (\p\overline{w}^+/\p x^+)$ that is the leading order spanwise pressure gradient term.}
    \label{fig:xzPotentialFlow}
\end{figure*}

The mechanism of transverse flow generation is elucidated by supporting the measurements with a theoretical flow model.
In this model, the flow is described as a uniform streamwise base flow that is perturbed by the groove geometry. 
Similar to Prandtl's original boundary-layer concept, and in line with the observations made in \S\ref{sec:flow-organisation-xy}, we may consider the flow disturbance as composed of an inertial (pressure-driven) \textit{`outer solution'} that is generated by the displacement effect of the non-smooth surface geometry, and a viscous \textit{`inner solution'} to accommodate the no-slip condition at the wall, which generates a near-wall spatial Stokes layer. 
The outer solution is modelled as a (steady, inviscid, and incompressible) potential flow that is subject to the flow-tangency condition at the wall. 
In case of a sufficient scale separation, this outer solution provides a good representation of the characteristics of the forcing, such as the maximum transverse velocity and the penetration height. 
The potential-flow model is described in detail in \ref{app:potential}, and relevant results will be referenced in the main text discussion where appropriate.

Figures~\ref{fig:xzPotentialFlow}(a,b) compare the results for the spanwise velocity component $\overline{w}$ in the wall-parallel plane, for the measurements and the potential-flow model (equation \eqref{eq:wPotential}).
For the measurements in figure~\ref{fig:xzPotentialFlow}(a), the streamlines within the groove show that the flow is deflected to align with the groove orientation, to form the primary flow structure established in \S\ref{sec:flow-organisation-xz}.
Instead, near the groove edges and where the groove slope is steep (around $x/\lambda_x = 0.25, 0.75$), a converging-diverging flow pattern appears, which causes the secondary flow on the flat wall between the grooves.
This converging-diverging flow is centred around the locations of maximum spanwise displacement of the grooves, i.e., $x/\lambda_x = 0, 0.5, 1$.
Upstream of the location at $x/\lambda_x = 0.5$, the flow is seen to enter the groove across its edge and converges, followed downstream by a diverging flow exiting the groove.
The same converging-diverging flow has been observed for dimples \citep{Nesselrooij_drag_2016}, which is understandable in view of the similarity between the local shape of the groove and the dimple geometry.

Figure~\ref{fig:xzPotentialFlow}(b) shows that potential flow provides a good prediction of the primary and secondary flow regions. 
The magnitudes are higher for the potential flow, which predicts the maximum spanwise velocity peak by evaluating the model at $y=0$ (elaborated in \S\ref{sec:forcing-characteristics}), whereas the PIV measurements reflect the (averaged) velocity further from the wall, which yields a lower $\overline{w}$, as can also be appreciated when comparing figures~\ref{fig:xzMeanFlow} and \ref{fig:xyMeanFlow}.
In the potential-flow model, the magnitudes of the primary and secondary flows are equal owing to the sinusoidal spanwise groove profile assumed (refer to \ref{app:potential}), i.e., the flat wall between grooves is represented by a sinusoidal ridge.
The converging-diverging flow topology is also well predicted.
These results provide confidence that potential flow can be used for first-order predictions of the flow over shallow surface deformations. 

Spanwise flow generation is ultimately driven through a bulk-forcing by the spanwise pressure gradient force $P_z$ induced by the SU test surface, which is governed by the spanwise momentum balance
\begin{equation}
    \label{eq:dpdz}
    P_z = -\frac{1}{\rho}\frac{\p \overline{p}}{\p z} = \overline{u_i }\frac{\p \overline{w}\ }{\p x_i} + \frac{\p \overline{w'u_i'}}{\p x_i} -\nu \frac{\p^2 \overline{w}}{\p x_i ^2},
\end{equation}
where $\rho$ is the fluid density ($i$ is the free index that denotes summing over the three coordinate directions).
An order of magnitude analysis, based on both experimental data and the potential-flow model, reveals that the streamwise advection (inertia) term constitutes the leading-order contribution, such that 
\begin{equation}
        P_z \approx \overline{u}\frac{\p \overline{w}}{\p x}.
\end{equation}
Figures~\ref{fig:xzPotentialFlow}(c,d) compare the measurements with potential flow for this leading-order term.
Despite noise in the measurements due to the spatial gradient computation, the overall organisation of $\overline{u}\p \overline{w}/\p x$ shows good correspondence with the potential-flow prediction. 
\begin{figure*}
    \centering
    \includegraphics[width = \textwidth]{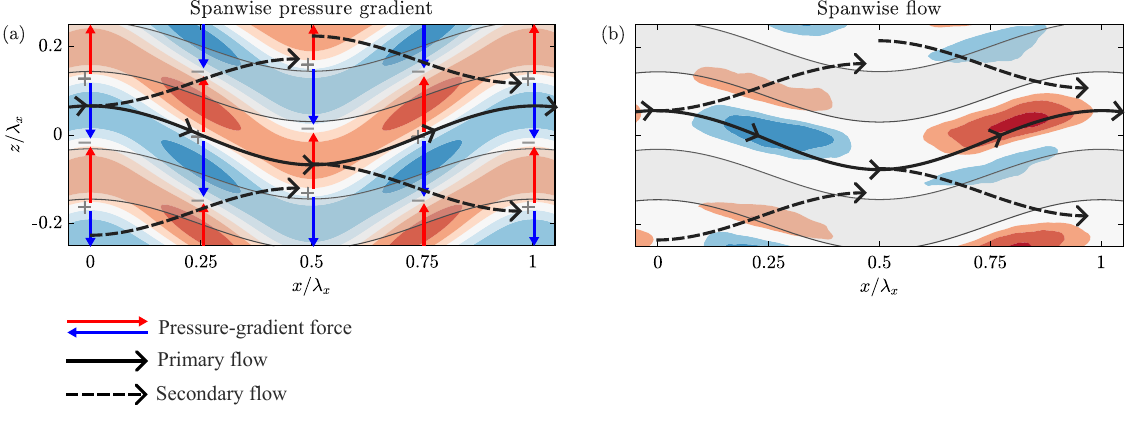}
    \caption{Conceptual sketch for the mechanism of spanwise flow generation. (a) Spanwise pressure gradient, where grey `$+$' and `$-$' indicate high- and low-pressure regions and red/blue arrows show the resultant pressure-gradient force.
    (b) mean spanwise flow induced by the SU test surface. Overlayed lines show the primary flow (solid) and secondary flow (dashed) generated.}
    \label{fig:PzSchematic}
\end{figure*}

The discussion regarding the working mechanism is supported by a conceptual sketch in figure~\ref{fig:PzSchematic}. 
Along the span, the maxima and minima in $P_z$ align with the centre of the grooves at $x/\lambda_x = 0, 0.5, 1$, i.e., the locations of the maximum lateral displacement of the groove baseline.
As a result, a streamwise altering pressure-gradient force pattern is established that 
\textit{`guides'} the flow along the groove axis. 
The spanwise amplitude of the $P_z$ iso-lines is larger than the spanwise displacement of the groove $A$, so that the $P_z$ peak inside the groove at $x/\lambda_x = 0$ moves to the flat-wall in between the grooves at $x/\lambda_x = 0.5$.
This causes a pressure gradient pattern near $x/\lambda_x = 0.25,0.75$ between the grooves that opposes the upstream primary $P_z$ inside the grooves, driving the secondary flow and ultimately resulting in the observed diverging-converging flow topology.

\section{Quantification of spanwise-forcing characteristics}\label{sec:forcing-characteristics}
The spanwise forcing effect is assessed by characterising the passive Stokes layer (PSL) induced by the SU test surfaces in terms of maximum spanwise velocity amplitude $W_{\max}$ and the outer-layer penetration depth $\Delta$. In \S\ref{sec:forcing-characteristics-geometry}, the geometric scaling of these properties is first presented using the potential-flow model in \ref{app:potential}
and subsequently compared to the experimental results in \S\ref{sec:forcing-characteristics-results}.

\subsection{Geometric scaling based on potential-flow model} \label{sec:forcing-characteristics-geometry}

The geometric scaling of $W_{\max}$ in terms of the SU geometric parameters follows from the potential-flow model according to equation \eqref{eq:wPotential}, as
\begin{equation}
    \label{eq:Wp}
    W_\text{pot} = Adk_x k_z U_c.
\end{equation}
As a first observation, we can derive from this expression the forcing efficiency with respect to the idealised 1D model, equation \eqref{eq:1Dmodel}, to be
\begin{equation}
    \label{eq:Wmax-efficiency}
   \frac{W_\text{pot}}{W_{0,1D}}  = dk_z = 2 \pi \frac{d}{\lambda_z} = 2 \pi \frac{D}{\lambda_z}\frac{d}{D}.
\end{equation}
Here, the groove width $D$ has been introduced as a convenient characteristic spanwise length scale to replace the spanwise wavelength $\lambda_z$; these parameters are in fixed proportion for all test surfaces considered ($\lambda_z/D = 1.65$, see Table~\ref{tab:modelGeometry}).
This shows that the forcing efficiency is predicted to be proportional to the depth-to-width ratio of the groove, and for the test surfaces with $d/D = 0.05$ this yields a theoretical forcing efficiency of 19\%, which is in excellent agreement with the observations in \S\ref{sec:mechanism}.

The scaling for the maximum spanwise velocity, equation \eqref{eq:Wp}, can be rewritten as 
\begin{equation}
    \label{eq:potentialScaling}
    \frac{
    W_\text{pot}}{U_c} = Adk_x k_z = c \frac{A}{\lambda_x} \frac{d}{D},
\end{equation}
where $c$ is a constant given by
\begin{equation}
    \label{eq:potentialScaling2}
       c = 2 \pi k_z D  = 4 \pi^2 \frac{D}{\lambda_z} \approx 24.
\end{equation}
Equation \eqref{eq:potentialScaling} predicts that the spanwise flow is (i) proportional to the convection velocity $U_c$, (ii) increases with increasing groove amplitude, where $A/\lambda_x$ determines the maximum groove slope and can be considered as the \textit{`groove-baseline aspect ratio'}, and (iii) increases (decreases) for deeper (shallower) groove depths, where $d/D$ can be considered as the \textit{`groove-profile aspect ratio'}.

Similarly, the scaling of the outer-layer penetration depth $\Delta$ follows from the the potential-flow model, equation \eqref{eq:Delta}, as
\begin{equation}
    \label{eq:Deltascaled}
    \frac{\Delta}{D} = \frac{1}{Dk_z} = \frac{\lambda_z}{2 \pi D} \approx 0.26.
\end{equation}
This reveals that $\Delta$ is predicted to follow a geometrical scaling and, to first order (i.e., within the approximations underlying the potential-flow model), to scale directly proportional to the groove width $D$, without further influence of the other dimensional characteristics, such as the groove depth ($d$), groove amplitude ($A$) and the groove streamwise length scale ($\lambda_x$).

\begin{figure}[t]
    \centering
    \includegraphics[width = \columnwidth]{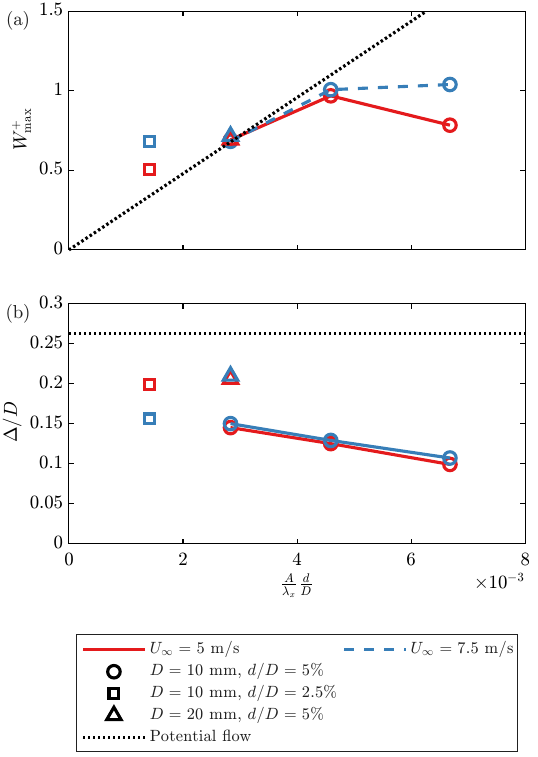}
    \caption{Quantification of the spanwise forcing parameters based on all measurements. (a) Maximum spanwise velocity $W^+_{\max}$ and (b) outer-layer penetration depth $\Delta/D$. Black dotted line gives potential flow estimates (a) $W_\text{pot} = Ad k_x k_z U_c$ for $U_c^+ = 10$, and (b) $\Delta = 1/k_z$.}
    \label{fig:forcingCharacteristics}
\end{figure}

\subsection{Passive Stokes layer characteristics}\label{sec:forcing-characteristics-results}
Figure~\ref{fig:forcingCharacteristics} displays the spanwise-forcing characteristics for all five test surfaces and for both free-stream velocities at $Re_\tau = 1020$ and 1410, shown in red (solid lines) and blue (dotted lines), respectively.
These characteristics are computed from the PIV data as outlined in \S\ref{sec:flow-organisation-xy}.
The scaling of the horizontal axis follows the suggestion of the potential-flow model, as discussed in \S\ref{sec:forcing-characteristics-geometry}.
Model A3-D10-d050 is considered as the baseline geometry, with $A/\lambda_x = 0.06$, $d/D=5\%$ and $D = 10$\:mm.  
The other test surfaces varied only one geometric parameter at a time: increasing groove amplitude (circle markers), decreasing groove depth to $d/D = 2.5\%$ (square markers), and scaling the self-similar groove geometry by a factor of two to $ D=20$ mm (triangular markers). 
For further details, see tables~\ref{tab:modelGeometry} and \ref{tab:modelGeometryViscous}.

As the groove amplitude $A$ increases (circle markers), $W_{\max}$ in figure~\ref{fig:forcingCharacteristics}(a) initially increases, in good agreement with the potential-flow scaling (dotted line), after which it levels and deviates from the linear scaling for the largest amplitude. 
This trend is consistent with the qualitative flow organisation observed in figure~\ref{fig:zAmplitudeTopology}.
Scaling of the velocity amplitude with $(A/\lambda_x)(d/D)$ is confirmed by the upscaled but self-similar groove (triangle markers), which exhibits the same $W_{\max}$ as the baseline groove (circle markers).
The shallower grooves (square markers) show a reduction in $W_{\max}$ of about half, and although the spread is large owing to experimental uncertainty, this trend aligns with the predicted theoretical scaling.
No strong influence of the free-stream velocity is observed as long as $(A/\lambda_x)(d/D)$ remains below $ 4.5\num{-3}$, confirming that the convection velocity in $W_{pot}$ and $W_{\max}$ scales proportionally with  $U_\tau$ (or $U_\infty$, the influence of $Re_\tau$ is small and these velocity scales are essentially linearly dependent). 
The highest velocity amplitude induced by the SU models is $W_{\max}^+ = 1$ at $A/\lambda_x = 0.09$ ($\frac{A}{\lambda_x} \frac{d}{D}= 4.5\num{-3}$), which amounts to approximately 5\% of $U_\infty$.

The potential-flow scaling breaks down when $\frac{A}{\lambda_x} \frac{d}{D}\gtrsim 4.5\num{-3}$, where the surface slope presumably becomes too steep (note that according to equation \eqref{eq:hx}, the surface gradient $\max(\p{h}/ \p{x})$ scales similarly as equation \eqref{eq:potentialScaling}), and flow separation may occur.
In this regime, the influence of $U_\infty$ is also apparent, indicating that the flow is indeed more complex and cannot be reliably predicted with a potential-flow model. 
Based on these results, $\frac{A}{\lambda_x} \frac{d}{D}\lesssim 4.5\num{-3}$, or equivalent $Adk_xk_z \lesssim 0.1$, is proposed as the limit on the maximum surface slope for the SU geometry.

Figure~\ref{fig:forcingCharacteristics}(b) shows the variation of $\Delta/D$, the outer-layer penetration depth normalised by the groove width. 
The experimental data appear to conform to a geometrical scaling of $\Delta$, but with deviations from the theoretical prediction of a constant value $\Delta/D$, although the order of magnitude agrees.  
A linearly decreasing trend between $\Delta/D$ and $(A/\lambda_x)(d/D)$ is observed. 
This trend does not break down for increasing $A/\lambda_x$ (circled markers), so it seems that the outer layer, produced by wall-normal decay of near-surface perturbations, is less susceptible to steep surface curvature than $W_{\max}$.
Its proportionality to $D$ is more complex; namely, the upscaled self-similar groove (triangle markers) has a higher $\Delta/D$ than the baseline model (circle markers). 
We can, however, confirm that the variation in $\Delta/D$ with respect to the model does not impact the prediction of $W_{\max}$ in view of the agreement between model and experiment in figure~\ref{fig:forcingCharacteristics}(a).

\section{Discussion on drag reduction by passive forcing}\label{sec:DR-discussion}
So far, we have considered only the inertial outer-layer solution for the passive Stokes layer (PSL), based on the potential-flow model, which enabled us to quantitatively relate the induced flow characteristics to the surface geometry. In this section, we use an analytical model of the full PSL solution that includes viscous effects, providing the complete inner-outer-layer structure established in \S\ref{sec:flow-organisation-xy}; its derivation is found in \ref{app:PSL}.
We first compare the PSL to the active SSL and experimental results in \S\ref{sec:DR-discussion-3D-model}.
To provide insight into the potential of the lateral flow forcing of the PSL, in \S\ref{sec:DR-discussion-active-comparison}, we match the forcing amplitude $W_{\max}$ to that of an equivalent SSL amplitude $W_{eq}$ (based on equivalence of the surface-averaged Stokes strain), so as to compare to the DR results in \citet{viotti_streamwise_2009}.

\subsection{Analytical model for the passive Stokes layer}\label{sec:DR-discussion-3D-model}
\begin{figure*}[t]
    \centering
    \includegraphics[width = \textwidth]{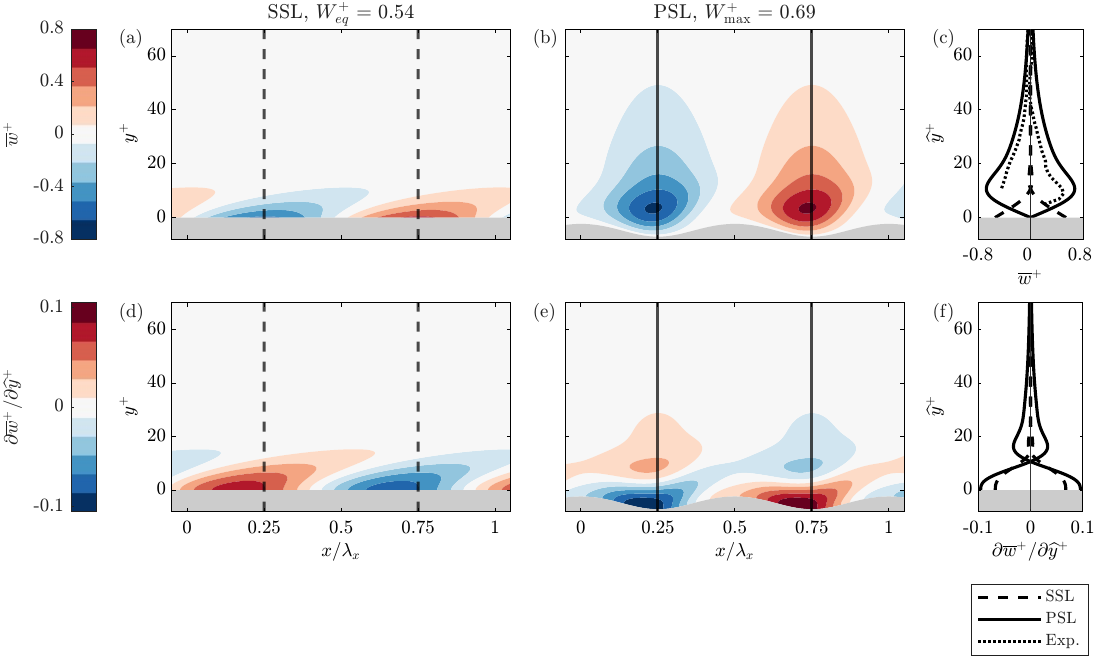}
    \caption{Comparison of the active spatial Stokes layer (SSL) and passive Stokes layer (PSL) for model A3-D10-d050.
    The first and second columns show streamwise-wall-normal contours for (a,d) the SSL, (b,e) the PSL; the third column (c,f) compares the wall-normal profiles at $x/\lambda_x = [0.25,0.75]$.
    The top and bottom row show (a-c) mean spanwise velocity, and (d-f) wall-normal Stokes strain $\partial \overline{w}^+ /\partial\widehat y^+$.}
    \label{fig:compareSSL}
\end{figure*}

A comparison between the experiments and theory is made for the baseline model A3-D10-d050 at $Re_\tau = 1020$.
The first and second columns of figure~\ref{fig:compareSSL} show the SSL and PSL analytical solutions, which are compared in the third column using wall-normal profiles. 
The experimental results as documented in \S\ref{sec:flow-organisation-xy} are included as well, and its characteristics $W_{\max}$ and $\Delta$ (given in figure~\ref{fig:forcingCharacteristics}) are used as input for the PSL model. 
The theoretical models (SSL and PSL) are evaluated at $z=0$; the top row shows the spanwise velocity $\overline{w}$, and the bottom row shows the Stokes strain $\p \overline{w}/\p y$.

In figure~\ref{fig:compareSSL}(a), the classic SSL solution is depicted, which is dominated by viscous effects as characterised by its near-wall ($y^+<15$), forward-inclined shear layer that is driven from the wall. 
Instead, the PSL in figure~\ref{fig:compareSSL}(b) conforms to the two-region structure introduced in \S\ref{sec:flow-organisation-xy}, characterised by a viscous inner layer below $y^+ <10$ to satisfy the no-slip condition, and an inertial (pressure-driven) outer layer that extends to a much higher wall-normal distance of $y^+ \approx 60$. 
Qualitatively, the $\overline{w}$ contours of the PSL match the experimental results in figure~\ref{fig:xyMeanFlow}(b), which is further supported by the comparison of the wall-normal profiles in figure~\ref{fig:compareSSL}(c).
There are some quantitative differences, for example, in the precise streamwise locations where $W_{\max}$ occurs and in the slightly stronger outer-layer solution of the analytical model. 
However, given the approximations underlying the theoretical model (refer to \ref{app:PSL}), the quantitative discrepancies are of secondary importance. The key result is the strong qualitative agreement between the PSL model and the experiment, which lends mutual support to the proposed flow organisation.

\subsection{Analogy with active forcing and drag reduction potential}\label{sec:DR-discussion-active-comparison}
In view of the essential differences in flow structure between the bulk-forced PSL and the actively wall-driven SSL, the velocity amplitude $W_{\max}$ of the PSL does not directly compare to that of the SSL, which prevents a one-on-one comparison to the $\text{DR}(W_{\text{SSL}})$ relation established in \citet{viotti_streamwise_2009}.
To overcome this discrepancy, $W_{\max}$ is mapped onto an equivalent active forcing amplitude $W_{eq}$, by matching the Stokes strain of the two models, similar to the approach taken in \citet{ghebali_large-scale_2017} for oblique wavy walls.
The Stokes strain is an appropriate choice for this procedure, since it, or derived measures, provides an effective diagnostic of DR \citep{choi_drag_2002, quadrio_critical_2004, yakeno2014modification, ding2023acceleration}, and it has proven relevant to the physical DR mechanism \citep{Touber_near-wall_2012, Agostini_spanwise_2014,agostini_turbulence_2015, knoop2025response}.
We consider the surface-averaged Stokes strain magnitude at the wall, computed as
\begin{equation}
    \mathcal{S} = \frac{1}{\lambda_x \lambda_z}\int_0^{\lambda_z}\int_0^{\lambda_x} \frac{\p|w|(\widehat y=0)}{\p y}\, dx\, dz.
\end{equation}
$\mathcal{S}$ is first computed for both models at unit amplitude, indicated by their respective subscripts. $W_{eq}$ is then related to $W_{\max}$ according to
\begin{equation}
    \label{eq:Weq}
    W_{eq} = \frac{\mathcal{S}_\text{PSL}}{\mathcal{S}_\text{SSL}} W_{\max}.
\end{equation}

The two models in figure~\ref{fig:compareSSL} are matched based on $\mathcal{S}$, such that $W^+_{\max} = 0.69$ (as observed in the experiment) corresponds to an effective amplitude of $W^+_{eq} = 0.54$. 
This lower effective amplitude is in part attributed to the spanwise-periodic PSL solution compared to the spanwise-uniform SSL solution, which would, theoretically, be a factor $2/\pi \approx 0.63$ lower given the average value of $|\cos(k_z z)|$. 
Consequently, the local near-wall Stokes strain in figures~\ref{fig:compareSSL}(d-f) is higher in the case of the PSL.
In practice, we find that the ratio is approximately 0.8, i.e., $W_{eq}\approx 0.8W_{\max}$, indicating that the PSL is less effective than the PSL at the same velocity amplitude.

Applying the same matching procedure for the strongest passive forcing case at $W^+_{\max} = 0.97$ (refer to figure~\ref{fig:forcingCharacteristics}), yields $W_{eq}^+ = 0.78$. 
To compare $W_{eq}^+ = 0.78$ to the DR reported in \citet{viotti_streamwise_2009}, we extrapolate their results at $\lambda_x = 1250$ at $Re_\tau = 200$ (these are effectively linear for their $W_\text{SSL}^+ =[1,2,6]$), for which we find an approximate frictional DR of 1.8\%.
This result suggests that the PSL produces sufficient Stokes strain to act on the spanwise-forcing drag-reduction mechanisms. 
Nevertheless, this small theoretical estimate is likely degraded by an unfavourable $Re_\tau$ scaling \citep{gatti_reynolds-number_2016} and added pressure drag due to the surface geometry.

To place this frictional DR estimate in context, for oblique wavy walls, a 1 to 3\% skin-friction DR was reported by \citet{ghebali_large-scale_2017}, which was largely offset by pressure drag, leading to either a small total DR of up to 0.7\% or an overall drag increase.
A similar conclusion was drawn for dimples in \citet{vanCampenhout2023experimental}. 
Using numerical simulations, they found a total drag increase of 1 to 2\%, whereas the experiments in \citet{Nesselrooij_drag_2016} initially reported a 4\% DR.

While a rigorous experimental characterisation of the total drag for all SU test surfaces is outside the scope of the present study, preliminary direct-force drag measurements of model A3-D10-d050 were conducted by \citet{Knoop2023spanwise}.
These results are mentioned here, not to draw quantitative conclusions, but rather to provide an impression in support of this discussion of the DR potential.
The measurements employed the direct force balance and methodology of \citet{van2022development}, which improved on that of \citet{Nesselrooij_drag_2016}, and were validated for dimples in \citet{vanCampenhout2023experimental} and riblets.
For SU model A3-D10-d050, as well as several other passive test surfaces, these measurements reported no change in total drag relative to a flat-plate reference within a $\pm 0.5\%$ uncertainty interval.

All these results suggest that, in line with earlier results on dimples and oblique wavy walls, the investigated SU geometry may produce a small amount of turbulent skin-friction DR, but this saving is likely offset by an increase in pressure drag, yielding a few per cent of net DR at most.
Any remaining DR potential will further degrade when implemented in industrial applications (e.g., $Re_\tau$-scaling and limited coverage), making it unlikely that passive spanwise forcing for turbulent DR will be practically relevant.

\section{Summary and conclusions}\label{sec:conclusions}
Sinusoidal surface grooves, meandering along the streamwise direction, and referred to as sinusoidal undulations (SU), were explored for their potential to passively generate transverse forcing, motivated by the high DR achievable via active spanwise forcing. 
Particle image velocimetry (PIV) in combination with analytical modelling was used to characterise the flow over five SU test surfaces, to elucidate the mechanisms of spanwise flow generation and the formation of the passive Stokes layer, and to support a discussion of its DR potential.

Results show that the SU test surfaces are effective at generating a streamwise-alternating transversal flow.
A primary flow is directed along the grooves, which is alternated in the spanwise direction by a secondary flow in opposing directions to produce a flow out of the groove across its edges; combined, this creates a converging-diverging flow pattern.
The mechanism of transverse flow generation is a spanwise pressure gradient $P_z$ induced by the surface. 
$P_z$ is driven by inertial effects caused by the surface displacement and the flow tangency condition, and produces an outer-layer solution that is modelled using potential flow. 
The potential-flow model represents this outer layer effectively and shows excellent qualitative and quantitative agreement with the experimental results.
Near the surface, viscous effects dominate, forming a thin inner layer to accommodate the no-slip condition. 
The combined inner-outer solution forms the \textit{passive Stokes layer} (PSL), for which an analytical solution is derived based on the pressure-gradient forced advection-diffusion equation of the spanwise momentum.

The potential-flow solution provided a theoretical estimate of the maximum spanwise velocity amplitude and its scaling with respect to the surface geometry.
When the spanwise cross-section remains self-similar (i.e., constant $d/D$ and $D/\lambda_z$), the predicted amplitude scales as: $W_\text{pot}\propto U_c \frac{A}{\lambda_x} \frac{d}{D}$, hence, being proportional to the convection velocity (taken as $U_c = 10 U_\tau$), to $A/\lambda_x$ signifying the aspect ratio between groove spanwise displacement and streamwise wavelength, and to the relative groove depth $d/D$. 
Experimental results confirm the validity of scaling the PSL characteristics using $\frac{A}{\lambda_x} \frac{d}{D}$, showing an excellent quantitative prediction of $W_{\max}^+ = 0.7$ for the baseline SU test-surface. 
Further confirmation of this scaling is provided by an upscaled but self-similar geometry (i.e., the same $\frac{A}{\lambda_x} \frac{d}{D} = 2.8\num{-3}$) that exhibits the same measured $W_{\max}$ as the baseline.
The effect of increasing groove amplitude $A/\lambda_x$ is initially an expected increase in velocity amplitude up to a maximum of $W_{\max}^+ = 1$, after which the velocity amplitude saturates and departs from the theoretical prediction.
This deviation is likely owing to the linearised boundary condition of the model and possible flow separation when the surface inclination becomes too steep.
We find that the theoretical scaling is appropriate when $\frac{A}{\lambda_x} \frac{d}{D}\lesssim  4.5\num{-3}$ or equivalently $A d k_x k_z \lesssim 0.1$.
The efficiency of the spanwise flow generation, compared to an initial assumption that the flow follows the groove perfectly, translating the streamwise velocity into a spanwise component, is approximately 20\% for both experiment and the theoretical prediction, and is found to depend on the spanwise cross-section $d k_z$ (related to $d/D$). 

The theoretical analyis of comparing the active and passive Stokes layers analytical solutions shows that the SU geometry is effective at inducing a modest Stokes strain at the surface, which suggests that this technique can act upon the established spanwise-forcing mechanism for skin-friction DR. Quantitatively matching the two models based on surface-averaged Stokes strain shows that $W_{\max}$ of the PSL translates into an approximately 20\% lower equivalent active wall-velocity amplitude, i.e., an efficiency of about 80\% as $W_{eq} \approx 0.8 W_{\max}$. 
While the real physics of how the PSL affects the flow turbulence may be more complex, this exercise provides an indication of its performance relative to the established DR characteristics of its active counterpart. Extrapolating the results of \citet{viotti_streamwise_2009} suggests an approximate 1.8\% reduction of frictional drag. 
These potential savings, however, are likely offset by additional pressure drag and other inefficiencies in practice.  
So, while frictional DR may be achieved with passive forcing, its potential as an alternative means to active spanwise forcing in practical applications appears to remain limited.

\section*{Acknowledgements}
The authors gratefully acknowledge the financial support of the Netherlands
Enterprise Agency under grant number TSH21002.
We are sincerely thankful to the team at Dimple Aerospace B.V., namely, Michiel van Nesselrooij, Olaf W.\,G. van Campenhout, and Friso H. Hartog, for their close collaboration and years of extensive research into passive flow control techniques aimed at turbulent skin-friction drag reduction.

\section*{Author's contributions}
\textbf{Max W. Knoop:} Conceptualisation, Methodology, Formal analysis, Investigation, Writing -- Original Draft, Writing  -- Review \& Editing, Visualisation, Supervision;
\textbf{Bas W. van Oudheusden:} Conceptualisation, Methodology, Writing -- Original Draft, Writing  -- Review \& Editing, Supervision;
\textbf{Luuk Pelkmans:} Methodology, Formal analysis, Investigation, Writing  -- Review \& Editing;
\textbf{Ferry F.\,J. Schrijer:} Methodology, Supervision.

\appendix

\section{Potential-flow model based on the surface geometry}
\label{app:potential}
As discussed in the main text (\S\ref{sec:mechanism}), a model for transverse flow actuation by SU surface grooves is proposed, based on a potential flow approach. For the convenience of the analysis, the surface geometry of streamwise aligned surface grooves is approximated by a spanwise sinusoidal groove profile, which gives a parametric description of the surface height
\begin{equation}
    \label{eq:potential-surfaceHeight}
    h(x,z) = -d\cos(k_z \hat z),\quad \text{where} \quad \hat z = z-A\cos(k_x x).
\end{equation}
Here $A$, $d$, $k_x$ and $k_z$ correspond to the geometric parameters introduced in \S\ref{sec:method-testsurfaces}.

The velocity vector $\mathbf{u} = (u,v,w)$ is expressed as the gradient of the velocity potential $\phi$:
\begin{equation}
    \label{eq:potential-U}
    \mathbf{u} = \nabla \phi.  
\end{equation}
The potential is decomposed into a streamwise-uniform flow at convection velocity $U_c$ and a small perturbation $\varphi$ induced by the surface geometry, as 
\begin{equation}
    \label{eq:potential}
    \phi(x,y,z) = U_c x + \varphi(x,y,z),
\end{equation}
The (perturbation) potential needs to satisfy the Laplace equation (mass conservation), i.e 
\begin{equation}
    \label{eq:potential-Laplace}
    \nabla^2 \phi =\nabla^2 \varphi = 0.
\end{equation}
with boundary conditions:
\begin{equation}
    \label{eq:potential-BC1}
    \varphi(x,\infty,0) = 0,
\end{equation}
\begin{equation}
    \label{eq:potential-BC2}
    \nabla \phi(x,0,z)\cdot \mathbf{n} = 0,
\end{equation}
where $\mathbf{n}$ is the surface normal vector given by
\begin{equation}
    \mathbf{n} = \left(-\frac{\p h}{\p x}, 1, -\frac{\p h}{\p z}\right).
\end{equation}
The conditions expressed in \eqref{eq:potential-BC1} and \eqref{eq:potential-BC2} correspond to, respectively, a decay of the perturbation as $y\to \infty$ and flow-tangency at the surface.

Using separation of variables, the solution of $\varphi$ is sought in the form 
\begin{equation}
    \label{eq:potential-pertubation}
    \varphi = g(x,z)\, f(y),
\end{equation}
Evaluating the Laplace equation, we find
\begin{equation}
    \label{eq:potential-Laplace2}
    \frac{f_{yy}}{f} = -\frac{g_{xx} + g_{zz}}{g}  = \text{constant}  
\end{equation}
where the subscripts denote partial differentiation with respect to the given variable. 
Equation \eqref{eq:potential-Laplace2} is constant (defined below) as a consequence of the variable separation.
Solving the left-hand side for $f$, with normalization $f(0) = 1$, yields

\begin{equation}
    \label{eq:f-solution}
    f(y) = e^{-y/\Delta},
\end{equation}
which reveals an exponential decay of the velocity perturbation, i.e., $f(\infty) = 0$, that satisfies the boundary condition in \eqref{eq:potential-BC1}. 
The exponential decay length, $\Delta$, introduced here is similar to the wall-normal penetration depth as defined in \S\ref{sec:flow-organisation-xy}. This result furthermore identifies the value of the constant in equation \eqref{eq:potential-Laplace2} as being equal to $1/{\Delta^2}$.

The second boundary condition in $\eqref{eq:potential-BC2}$, the flow-tangency condition at the surface, is linearised around $y=0$; neglecting the higher-order terms gives
\begin{equation}
    \frac{\p \varphi(x,0,z) }{\p y} = U_c \frac{\p h(x,z) }{\p x}.
\end{equation}
Substituting equations \eqref{eq:potential-pertubation} and \eqref{eq:f-solution} then yields 
\begin{equation}
    g(x,z) = -U_c \Delta \frac{\p h(x,z) }{\p x},
\end{equation}

The estimate of the penetration depth $\Delta$ is provided by evaluating the right-hand side of \eqref{eq:potential-Laplace2}.
Here, the streamwise gradient term $g_{xx}$ is neglected w.r.t. $g_{zz}$, since the characteristic spanwise length scale is much smaller than the streamwise one, i.e., $\lambda_x\gg \lambda_z$. With this approximation, we find 
\begin{equation}
    \frac{1}{\Delta^2} \approx -\frac{g_{zz}}{g} = -\frac{h_{xzz}}{h_x},
\end{equation}
where
\begin{equation}
    \label{eq:hx}
    h_x = \frac{\p h(x,z)}{\p x} = dA k_x k_z \sin(k_x x)\sin(k_x \widehat z),
\end{equation}
and
\begin{equation}
    \begin{aligned}
        h_{xzz} & =  \frac{\p^3 h(x,z)}{\p x \p z^2} 
         = -dA k_x k_z^3 \sin(k_x x)\sin(k_x \widehat z)       
    \end{aligned}
\end{equation}
which gives
\begin{equation}
    \label{eq:Delta}
    \Delta = \frac{1}{k_z}.
\end{equation}

Collecting all results, the perturbation potential can be expressed as
\begin{equation}
    \begin{aligned}
      \varphi(x,y,z) & = -U_c \Delta \frac{\p h(x,z) }{\p x} e^{-y/\Delta} \\
      & =  -U_c A dk_x \sin(k_x x)\cos(k_z \hat z)e^{-y/\Delta}.
    \end{aligned}
\end{equation}
Solving for the velocity components yields

\begin{equation}
    u(x,y,z)  = U_c  + \frac{\p \varphi}{\p x} \approx U_c,
\end{equation}
\begin{equation}
    \label{eq:v0}
    \begin{aligned}
       v(x,y,z) & =  \frac{\p \varphi}{\p z} 
       \\ & = U_c A dk_x k_z \sin(k_x x)\cos(k_z \hat z)e^{-y/\Delta}
    \end{aligned}
\end{equation}
and 
\begin{equation}
    \label{eq:wPotential}
    \begin{aligned}
        w(x,y,z) & = \frac{\p \varphi}{\p z}   \\
          & =  -\underbrace{U_c A dk_x k_z  }_{W_\text{pot}}\sin(k_x x)\cos(k_z \hat z)e^{-y/\Delta}.
    \end{aligned}
\end{equation}
$W_\text{pot}$ provides the maximum spanwise velocity at $y=0$, which serves as a first-order estimate of the maximum spanwise velocity induced by the surface grooves.

\section{Three-dimensional model of the passive Stokes layer}
\label{app:PSL}
To complement the potential flow model derived in Appendix A, which represents the outer layer of the passive Stokes layer (PSL), this Appendix considers a viscous flow model for the entire PSL by including a viscous term in order to satisfy the no-slip condition at the wall. 
The analysis follows the same (laminar flow) approach, which \cite{viotti_streamwise_2009} applied to the active spatial Stokes layer (SSL). 
Here, the PSL is considered as a perturbation of a parallel base flow $(u(y),0,0)$. When further assuming that the thickness of the PSL is small with respect to the streamwise wavelength, the (linearised) spanwise-momentum equation reads:
\begin{equation}
    \begin{aligned}
    &u \frac{\p w}{\p x}  = \underbrace{-\frac{1}{\rho}\frac{\p p}{\p z}}_{P_z} + \nu \frac{\p^2 w}{\p y^2}.      
    \end{aligned}
\end{equation}
where $P_z$ represents the spanwise pressure-gradient force. 

When further assuming a linear streamwise velocity profile, the governing equation becomes
\begin{equation}
    \label{eq:3D-govEq1}
    (u_0y)\frac{\p w}{\p x} - \nu \frac{\p^2 w}{\p y^2}= P_z,     
\end{equation}
where $u_0$ is the wall-normal gradient of the streamwise velocity at the surface, i.e., $(\p u/\p y)_{y=0} = \nu U_\tau ^2 = U_\tau / \delta_\nu$.
This result is identical to the governing equation for the SSL, as considered by \citet{viotti_streamwise_2009}, apart from the additional forcing term $P_z$.
The SSL solution, hence, provides the homogeneous solution of \eqref{eq:3D-govEq1}, and the total solution can therefore be expressed as a combination with a particular solution arising from the pressure-gradient forcing.

We first simplify the three-dimensional solution into a planar solution, using the coordinate transformation $\widehat z = z - A\cos(k_z x)$ (previously introduced in \eqref{eq:potential-surfaceHeight}) that represents the spanwise location relative to the groove baseline, and by defining $w(x,y,z) = \widetilde w(x,y)\cos (k_z \widehat z)$ and $P_z(x,y,z) = \widetilde P_z(x,y)\cos (k_z \widehat z)$. 
Introducing these transformations in \eqref{eq:3D-govEq1} and dividing out the term $\cos (k_z \widehat z)$ yields the spanwise momentum equation for $\widetilde w(x,y)$, 
\begin{equation}
    \label{eq:3D-govEq2}
    \begin{aligned}
        & u_0y\frac{\p \widetilde w}{\p x} - \nu \frac{\p^2 \widetilde w}{\p y^2}=  \widetilde P_z.\\ 
    \end{aligned}
\end{equation} 

In view of \eqref{eq:3D-govEq2}, an estimate of the magnitude of $P_z$ is based on the consideration that in the outer layer the pressure gradient is balanced by the inertia term, such that
\begin{equation}
    \widetilde P_z \sim   -(u_0y)\frac{\p \widetilde w_o}{\p x}.     
\end{equation}
where $\widetilde w_o$ refers to the outer solution, as represented by equation \eqref{eq:wPotential}.
This leads to the following model for the spanwise pressure gradient:
\begin{equation}
    \label{eq:3D-PzModel}
    \widetilde P_z = -(u_0y) P_0 \cos(k_x x)e^{-y/\Delta},
\end{equation}
where $P_0$ is a constant, given by $P_0= A d k_x^2 k_z U_c$ according to the potential-flow model.
For the current purpose, however, $P_0$ is treated as a constant of which the value can be tuned, such as to produce a desired value of the velocity amplitude $W_{\max}$.

By introducing $\widetilde w(x,y) = \Re[e^{ik_x x}E(y)]$ into \eqref{eq:3D-govEq2}, an ordinary differential equation is obtained: 
\begin{equation}
    i \delta_s^{-3}y E - E_{yy} = -\frac{u_0 P_0}{\nu}\, y\, e^{-y/\Delta},
\end{equation}
with the Stokes layer thickness defined as
\begin{equation}
    \delta_s = \left( \frac{\nu}{u_0 k_x}\right )^{1/3}.
\end{equation}
Introducing the change of variable
\begin{equation}
    y = e^{-i\pi/6}\delta_s\xi,
    \quad \text{and}\quad
    F(\xi)=E(e^{-i\pi/6}\delta_s\xi),
\end{equation}
rotates the wall-normal coordinate into the decaying sector of the Airy equation and transforms the homogeneous problem into its canonical form, as given in \citet{viotti_streamwise_2009}.  
This choice ensures that the physically admissible solution is given by the decaying Airy function of the first kind, i.e., $\Ai(\xi)\to 0$ as $\xi\to\infty$. The governing equation then becomes
\begin{equation}
    \label{eq:PSL-forced-ODE}
   F_{\xi \xi}  - \xi F  = \underbrace{Q \xi e^{-\xi/\Delta_{\xi}}}_{f(\xi)}, 
\end{equation}
with
\begin{equation}
    Q = -\frac{iP_0}{k_x},\quad \text{and}\quad \Delta_\xi = \frac{\Delta}{e^{-i\pi/6} \delta_s}.
\end{equation}
The left-hand side of \eqref{eq:PSL-forced-ODE} corresponds to the Airy equation with linear operator $\mathcal{L} = (\mathrm{d}/\mathrm{d}\xi)^2 - \xi$ that is forced by $f(\xi)$ on the right-hand side.
The solution for $F(\xi)$ is sought in the form of its homogeneous and particular solution,
\begin{equation}
    F(\xi) = F_h(\xi) + F_p(\xi).
\end{equation}

The homogeneous equation admits two linearly independent solutions given by the Airy functions of the first and second kind, $\Ai(\xi)$ and $\Bi(\xi)$. Since $\Ai(\xi)$ is the only function that decays exponentially as $\xi\to \infty$, while $\Bi(\xi\to \infty)$ grows without bound, the solution to the homogeneous equation is $F_h(\xi) = \gamma \Ai(\xi)$. The full solution is then written as
\begin{equation}
    \begin{aligned}
        F(\xi) = \gamma \Ai(\xi) + F_p(\xi),
    \end{aligned}
\end{equation}
where the constant $\gamma$ is determined from the no-slip boundary condition $\widetilde w(x,0) = 0$ (equivalently $F(0) = 0$) and found to be
\begin{equation}
    \gamma = -\frac{F_p(0)}{\Ai(0)}.
\end{equation}

Green’s function is used to construct the particular solution. 
To this end, the two fundamental solutions $\Bi(\xi)$ and $\Ai(\xi)$ are chosen such that the first one satisfies the boundary condition at the wall i.e., $\Bi(0)\neq 0$, which is non-zero to ensure $F(0) = 0$ since $F_h(0)\neq 0$, and the second one satisfies decay in the far field, i.e., $\Ai (\xi \to \infty) = 0$. 
The Green’s function associated with the operator $\mathcal{L}$ of \eqref{eq:PSL-forced-ODE} and satisfying decay in the far field is given by \citep{morse1946methods}
\begin{equation}
   G(\xi,s) =  \frac{1}{W}
    \begin{cases}
        \Bi(s) \Ai(\xi), & s<\xi,\\
        \Bi(\xi) \Ai(s), & s>\xi,
    \end{cases}
\end{equation}
where the Wronskian of the fundamental two solutions is constant and given by
\begin{equation}
    W = \Bi(\xi) \Ai'(\xi) - \Bi'(\xi) \Ai(\xi) = \frac{1}{\pi}. 
\end{equation}
The particular solution is then obtained using the convolution of the Green’s function $G(\xi,s)$ with the forcing term:
\begin{equation}
    \begin{aligned}
        F_p(\xi) = & \int_0^\infty G(\xi,s) f(s) ds =\\
        &\pi Q \Ai(\xi) \int_0^\xi \Bi(s) s\, e^{-s/\Delta}  ds +\\
         &\pi Q \Bi(\xi) \int_\xi^\infty \Ai(s) s\,e^{-s/\Delta} ds.
    \end{aligned}
\end{equation}

Collecting all results, the analytical solution for the PSL is given by 
\begin{equation}
\begin{aligned}
   & w(x,y,z) = \cos(k_z \widehat z)\, \Re\Bigg[e^{ikx} \bigg\{ \gamma \Ai\! \left(e^{i\pi/6}\frac{y}{\delta_s}\right) +\\
   & \pi Q\, \Ai\! \left(e^{i\pi/6}\frac{y}{\delta_s}\right)\int_0^{e^{i\pi/6}y/\delta_s} \Bi(s) s\, e^{-s/\Delta_\xi}\,ds +\\
   & \pi Q\, \Bi\! \left(e^{i\pi/6}\frac{y}{\delta_s}\right)\int_{e^{i\pi/6}y/\delta_s}^{\infty}\Ai(s)s\, e^{-s/\Delta_\xi}\,ds \bigg\} \Bigg].
\end{aligned}
\end{equation}

\bibliographystyle{elsarticle-harv} 
\bibliography{2-bibliography}

@article{kim1993propagation,
  title={Propagation velocity of perturbations in turbulent channel flow},
  author={Kim, J. and Hussain, F.},
  journal={Phys. Fluids A: Fluid Dyn.},
  volume={5},
  number={3},
  pages={695--706},
  year={1993},
doi = {10.1063/1.858653},
  publisher={American Institute of Physics}
}

@article{choi_drag_2002,
	title = {Drag Reduction by Spanwise Wall Oscillation in Wall-Bounded Turbulent Flows},
	volume = {40},
	issn = {0001-1452},
	doi = {10.2514/2.1750},
	pages = {842--850},
	number = {5},
	journal = {AIAA J.},
	author = {Choi, J.-I. and Xu, C.-X. and Sung, H. J.},
	year = {2002},
}

@article{gatti_reynolds-number_2016,
	title = {Reynolds-number dependence of turbulent skin-friction drag reduction induced by spanwise forcing},
	volume = {802},
	issn = {0022-1120},
	doi = {10.1017/jfm.2016.485},
	pages = {553--582},
	journal = {J. Fluid Mech.},
	author = {Gatti, D. and Quadrio, M.},
	year = {2016},
}

@article{quadrio_critical_2004,
	title = {Critical assessment of turbulent drag reduction through spanwise wall oscillations},
        doi = {10.1017/S0022112004001855},
	volume = {521},
	pages = {251--271},
	journal = {J. Fluid Mech.},
	author = {Quadrio, M. and Ricco, P.},
	year = {2004},
}

@article{quadrio_laminar_2011,
	title = {The laminar generalized Stokes layer and turbulent drag reduction},
	volume = {667},
	issn = {0022-1120},
	doi = {10.1017/s0022112010004398},
	pages = {135--157},
	journal = {J. Fluid Mech.},
	author = {Quadrio, M. and Ricco, P.},
	year = {2011},
}

@article{quadrio_streamwise-travelling_2009,
	title = {Streamwise-travelling waves of spanwise wall velocity for turbulent drag reduction},
	volume = {627},
	issn = {1469-7645},
        doi = {10.1017/S0022112009006077},
	pages = {161--178},
	journal = {J. Fluid Mech.},
	author = {Quadrio, M. and Ricco, P. and Viotti, C.},
	year = {2009},
}

@article{ricco_effects_2004,
	title = {On the effects of lateral wall oscillations on a turbulent boundary layer},
	volume = {29},
	issn = {0894-1777},
	doi = {10.1016/j.expthermflusci.2004.01.010},
	pages = {41--52},
	number = {1},
	journal = { Exp. Therm. Fluid Sci.},
	author = {Ricco, P. and Wu, S.},
	year = {2004},
}

@article{ricco_review_2021,
	title = {A review of turbulent skin-friction drag reduction by near-wall transverse forcing},
	volume = {123},
	issn = {0376-0421},
        doi = {10.1016/j.paerosci.2021.100713},
	pages = {100713},
	journal = {Prog. Aerosp. Sci.},
	author = {Ricco, P and Skote, M and Leschziner, MA},
	year = {2021},
}

@article{viotti_streamwise_2009,
	title = {Streamwise oscillation of spanwise velocity at the wall of a channel for turbulent drag reduction},
	volume = {21},
    pages = {115109},
	doi = {10.1063/1.3266945},
	journal = {Phys. Fluids},
	author = {Viotti, C. and Quadrio, M. and Luchini, P.},
	year = {2009},
}

@article{ghebali_large-scale_2017,
   author = {Ghebali, S and Chernyshenko, SI and Leschziner, MA},
   title = {Can large-scale oblique undulations on a solid wall reduce the turbulent drag?},
   journal = {Phys. Fluids},
   volume = {29},
   number = {10},
   pages = {105102},
   ISSN = {1070-6631},
    doi = {10.1063/1.5003617},
   year = {2017},
   type = {Journal Article}
}

@article{Nesselrooij_drag_2016,
   author = {van Nesselrooij, Michiel and Veldhuis, LLM and Van Oudheusden, BW and Schrijer, FFJ},
   title = {Drag reduction by means of dimpled surfaces in turbulent boundary layers},
   journal = {Exp. Fluids},
   volume = {57},
   number = {9},
   pages = {1-14},
   ISSN = {1432-1114},
   year = {2016},
   type = {Journal Article},
doi = {10.1007/s00348-016-2230-9},
}

@article{ding2023acceleration,
  title={Acceleration is the key to drag reduction in turbulent flow},
  author={Ding, Liuyang and Sabidussi, Lena F and Holloway, Brian C and Hultmark, Marcus and Smits, Alexander J},
  journal={Proc. Natl. Acad. Sci.},
  volume={121},
  number={43},
  pages={e2403968121},
  year={2024},
DOI = {10.1073/pnas.2403968121},
  publisher={National Academy of Sciences}
}

@article{Agostini_spanwise_2014,
title={Spanwise oscillatory wall motion in channel flow: drag-reduction mechanisms inferred from DNS-predicted phase-wise property variations at $\textit{Re}_{\tau} =1000$}, 
volume={743}, 
DOI={10.1017/jfm.2014.40}, 
journal = {J. Fluid Mech.},
publisher={Cambridge University Press}, 
author={Agostini, L. and Touber, E. and Leschziner, M. A.}, 
year={2014}, 
pages={606–635}}

@article{Touber_near-wall_2012,
   author = {Touber, Emile and Leschziner, Michael A.},
   title = {Near-wall streak modification by spanwise oscillatory wall motion and drag-reduction mechanisms},
   journal = {J. Fluid Mech.},
   volume = {693},
   pages = {150-200},
   ISSN = {0022-1120},
   DOI = {10.1017/jfm.2011.507},
   year = {2012},
   type = {Journal Article}
}

@article{agostini_turbulence_2015,
  title={The turbulence vorticity as a window to the physics of friction-drag reduction by oscillatory wall motion},
  author={Agostini, L and Touber, E and Leschziner, MA},
  journal={Int. J. Heat Fluid Fl.},
  volume={51},
  pages={3--15},
  year={2015},
DOI = {https://doi.org/10.1016/j.ijheatfluidflow.2014.08.002},
  publisher={Elsevier}
}

@article{jimenez1999autonomous,
  title={The autonomous cycle of near-wall turbulence},
  author={Jim{\'e}nez, Javier and Pinelli, Alfredo},
  journal={J. of Fluid Mech.},
  volume={389},
  pages={335--359},
DOI = {10.1017/S0022112099005066},
  year={1999},
  publisher={Cambridge University Press}
}

@article{kline1967structure,
  title={The structure of turbulent boundary layers},
  author={Kline, Stephen J and Reynolds, William C and Schraub, Frederic Anthony and Runstadler, Peter W},
  journal={J. Fluid Mech.},
  volume={30},
  number={4},
  pages={741--773},
  year={1967},
  publisher={Cambridge University Press}
}

@misc{vanNesselrooij2020body,
  title={Body provided with a superficial area adapted to reduce drag},
  author={Van Nesselrooij, M and Van Campenhout, O and Veldhuis, LLM},
  year={2020},
  month=dec # "~1",
  publisher={Google Patents},
  note={US Patent 10,851,817}
}

@article{vanCampenhout2023experimental,
  title={Experimental and numerical investigation into the drag performance of dimpled surfaces in a turbulent boundary layer},
  author={Van Campenhout, OWG and Van Nesselrooij, M and Lin, YY and Casacuberta, J and van Oudheusden, BW and Hickel, S},
  journal={Int. J. Heat Fluid Flow},
  volume={100},
  pages={109110},
  year={2023},
  publisher={Elsevier}
}

@article{carrasco2024experimental,
  title={Experimental Investigation into the Drag Performance of Chevron-Shaped Protrusions in Wall-Bounded Turbulence},
  author={Carrasco Grau, Julio and van Campenhout, Olaf WG and Hartog, Friso H and van Nesselrooij, Michiel and Baars, Woutijn J and Schrijer, Ferdinand FJ},
  journal={Flow Turbul. Combust.},
  volume={113},
  number={1},
  pages={159--175},
  year={2024},
  publisher={Springer}
}

@article{auteri2010experimental,
  title={Experimental assessment of drag reduction by traveling waves in a turbulent pipe flow},
  author={Auteri, F. and Baron, A. and Belan, M. and Campanardi, G. and Quadrio, M.},
  journal={Phys. Fluids},
  volume={22},
  number={11},
  year={2010},
  publisher={AIP Publishing},
DOI = {10.1063/1.3491203}
}

@article{bird2018experimental,
  title={Experimental control of turbulent boundary layers with in-plane travelling waves},
  author={Bird, J. and Santer, M. and Morrison, J.F.},
  journal={Flow Turbul. Combust.},
  volume={100},
  pages={1015--1035},
  year={2018},
  publisher={Springer},
DOI = {10.1007/s10494-018-9926-2}
}

@article{gatti2015experimental,
  title={Experimental assessment of spanwise-oscillating dielectric electroactive surfaces for turbulent drag reduction in an air channel flow},
  author={Gatti, D. and G{\"u}ttler, A. and Frohnapfel, B. and Tropea, C.},
  journal={Exp. Fluids},
  volume={56},
  pages={1--15},
  year={2015},
  publisher={Springer},
DOI = {10.1007/s00348-015-1983-x}
}

@article{choi1998turbulent,
  title={Turbulent boundary-layer control by means of spanwise-wall oscillation},
  author={Choi, K.-S. and DeBisschop, J.-R. and Clayton, Brian R},
  journal={AIAA journal},
  volume={36},
  number={7},
  pages={1157--1163},
  year={1998}
}

@article{choi2001mechanism,
  title={The mechanism of turbulent drag reduction with wall oscillation},
  author={Choi, Kwing-So and Clayton, Brian R},
  journal={International Journal of Heat and Fluid Flow},
  volume={22},
  number={1},
  pages={1--9},
  year={2001},
  publisher={Elsevier}
}

@article{lienhart2008drag,
  title={Drag reduction by dimples?--A complementary experimental/numerical investigation},
  author={Lienhart, Hermann and Breuer, Michael and K{\"o}ksoy, Cagatay},
  journal={International Journal of Heat and Fluid Flow},
  volume={29},
  number={3},
  pages={783--791},
  year={2008},
  publisher={Elsevier}
}

@article{tay2015mechanics,
  title={Mechanics of drag reduction by shallow dimples in channel flow},
  author={Tay, C M J and Khoo, B C and Chew, Y T},
  journal={Phys. Fluids},
  volume={27},
  number={3},
  year={2015},
  publisher={AIP Publishing}
}

@article{chernyshenko2013drag,
  title={Drag reduction by a solid wall emulating spanwise oscillations. Part 1},
  author={Chernyshenko, Sergei},
  journal={arXiv:1304.4638},
  year={2013}
}

@article{scarano2000advances,
  title={Advances in iterative multigrid {PIV} image processing},
  author={Scarano, Fulvio and Riethmuller, Michel L},
  journal={Exp. Fluids},
  volume={29},
  number={Suppl 1},
  pages={S051--S060},
  year={2000},
  publisher={Springer}
}

@article{scarano2002iterative,
  title={Iterative image deformation methods in {PIV}},
  author={Scarano, Fulvio},
  journal={Meas. Sci. Technol.},
  volume={13},
  number={1},
  pages={R1--R19},
  year={2002}
}

@article{cafiero2024manipulation,
  title={Manipulation of a turbulent boundary layer using sinusoidal riblets},
  author={Cafiero, Gioacchino and Amico, Enrico and Iuso, Gaetano},
  journal={J. Fluid Mech.},
  volume={984},
  pages={A59},
  year={2024},
  publisher={Cambridge University Press}
}

@article{cafiero2022drag,
  title={Drag reduction in a turbulent boundary layer with sinusoidal riblets},
  author={Cafiero, Gioacchino and Iuso, Gaetano},
  journal={Exp. Therm. Fluid Sci.},
  volume={139},
  pages={110723},
  year={2022},
  publisher={Elsevier}
}

@article{sasamori2014experimental,
  title={Experimental study on drag-reduction effect due to sinusoidal riblets in turbulent channel flow},
  author={Sasamori, M and Mamori, H and Iwamoto, K and Murata, A},
  journal={Exp. Fluids},
  volume={55},
  number={10},
  pages={1828},
  year={2014},
  publisher={Springer}
}

@inproceedings{peet2008turbulent,
  title={Turbulent drag reduction using sinusoidal riblets with triangular cross-section},
  author={Peet, Yulia and Sagaut, Pierre and Charron, Yves},
  booktitle={38th Fluid Dynamics Conference and Exhibit},
  pages={3745},
  year={2008}
}

@article{knoop2025response,
  title={Response of a turbulent boundary layer to steady, square-wave-type transverse wall-forcing},
  author={Knoop, M.W. and Deshpande, R. and Schrijer, F. F. J. and van Oudheusden, B. W.},
  journal={Phys. Rev. Fluids},
  volume={10},
  number={6},
  pages={064607},
  year={2025},
DOI = {10.1103/PhysRevFluids.10.064607},
  publisher={APS}
}

@mastersthesis{Schaafsma2025transverse,
  author       = {Schaafsma, S H},
  title        = {Transverse Forcing by Acoustic Excitation},
  school       = {Delft University of Technology},
  year         = {2025},
  address      = {Delft, The Netherlands},
  type         = {Master's thesis},
  url         = {https://resolver.tudelft.nl/uuid:554fdd0c-ef09-4156-8fa9-756523fd8f7f}
}

@article{yakeno2014modification,
  title={Modification of quasi-streamwise vortical structure in a drag-reduced turbulent channel flow with spanwise wall oscillation},
  author={Yakeno, Aiko and Hasegawa, Yosuke and Kasagi, Nobuhide},
  journal={Phys. Fluids},
  volume={26},
  number={8},
  year={2014},
  pages = {085109},
  publisher={AIP Publishing}
}

@article{van2022development,
  title={Development of an experimental apparatus for flat plate drag measurements and considerations for such measurements},
  author={Van Nesselrooij, M and Van Campenhout, OWG and Van Oudheusden, BW and Schrijer, FFJ and Veldhuis, LLM},
  journal={Meas. Sci. Technol.},
  volume={33},
  number={5},
  pages={055303},
  year={2022},
  publisher={IOP Publishing}
}

@mastersthesis{Knoop2023spanwise,
  author       = {Knoop, M W},
  title        = {Spanwise forcing for turbulent drag reduction},
  school       = {Delft University of Technology},
  year         = {2023},
  address      = {Delft, The Netherlands},
  type         = {Master's thesis},
  url         = {https://resolver.tudelft.nl/uuid:b60d4b92-da7a-4fdb-bac4-2339f2105bdc}
}

@book{morse1946methods,
  title={Methods of theoretical physics},
  author={Morse, Philip McCord and Feshbach, Herman},
  year={1946},
  publisher={Technology Press}
}

@article{breuer2004actuation,
  title={Actuation and control of a turbulent channel flow using Lorentz forces},
  author={Breuer, Kenneth S and Park, Jinil and Henoch, Charles},
  journal={Phys. Fluids},
  volume={16},
  number={4},
  pages={897--907},
  year={2004},
  publisher={American Institute of Physics}
}

@article{berger2000turbulent,
  title={Turbulent boundary layer control utilizing the Lorentz force},
  author={Berger, Timothy W and Kim, John and Lee, Changhoon and Lim, Junwoo},
  journal={Phys. Fluids},
  volume={12},
  number={3},
  pages={631--649},
  year={2000},
  publisher={American Institute of Physics}
}

@article{corke2018active,
  title={Active and passive turbulent boundary-layer drag reduction},
  author={Corke, Thomas C and Thomas, Flint O},
  journal={AIAA J.},
  volume={56},
  number={10},
  pages={3835--3847},
  year={2018},
  publisher={American Institute of Aeronautics and Astronautics}
}

@article{lee2002control,
  title={Control of the viscous sublayer for drag reduction},
  author={Lee, Changhoon and Kim, John},
  journal={Phys. Fluids},
  volume={14},
  number={7},
  pages={2523--2529},
  year={2002},
  publisher={American Institute of Physics}
}

@article{thomas2019turbulent,
  title={Turbulent drag reduction using pulsed-DC plasma actuation},
  author={Thomas, FO and Corke, TC and Duong, A and Midya, S and Yates, K},
  journal={	J. Phys. D: Appl. Phys.},
  volume={52},
  number={43},
  pages={434001},
  year={2019},
  publisher={IOP Publishing}
}

@inproceedings{wilkinson2003investigation,
  title={Investigation of an oscillating surface plasma for turbulent drag reduction},
  author={Wilkinson, Stephen P},
  booktitle={41st Aerospace Sciences Meeting and Exhibit},
  number={AIAA Paper 2003-1023},
  year={2003},
  pages = {1023}
}

@article{launder1983turbulent,
  title={The turbulent wall jet measurements and modeling},
  author={Launder, BE and Rodi, W},
  journal={Annu. Rev. Fluid Mech.},
  volume={15},
  number={1},
  pages={429--459},
  year={1983},
  publisher={Annual Reviews 4139 El Camino Way, PO Box 10139, Palo Alto, CA 94303-0139, USA}
}





\end{document}